\documentclass{aa}

\usepackage{graphics}

\def\arcmin{\hbox{$^\prime$}}

\def\fm{\hbox{$.\!\!^{\rm m}$}}  
\def\fs{\hbox{$.\!\!^{\rm s}$}}  
\def\fdg{\hbox{$.\!\!^\circ$}}  
  
\def\farcs{\hbox{$.\!\!^{\prime\prime}$}}  
 
\def \hi {H\,{\sc i~}} 
 
\def \Mg {Mg\,{\sc ii~}}
\def\kms{km\,s$^{-1}$}

\begin{document} 
\thesaurus{10(09.03.1; 10.11.1; 11.01.1; 11.09.4; 11.12.1)} 
\title{High--resolution imaging of compact high--velocity clouds} 
\titlerunning{High--resolution imaging of CHVCs} 
\author{R. Braun\inst{1} and W. B. Burton\inst{2} } 
\institute{Netherlands Foundation for Research in Astronomy, P.O. Box 2, 
7990 AA Dwingeloo, The Netherlands 
\and Sterrewacht Leiden, P.O. Box 9513, 2300 RA Leiden, The Netherlands } 

\date{Received mmddyy/ accepted mmddyy} 
\offprints{R. Braun} 
\maketitle 

\begin{abstract} 
  
  Six examples of the compact, isolated \hi high--velocity clouds
  (CHVCs) identified by Braun and Burton (\cite{brau99}), but only
  marginally resolved in single--dish data, have been imaged with the
  Westerbork Synthesis Radio Telescope. The 65 confirmed objects in
  this class define a dynamically cold system, with a global minimum
  for the velocity dispersion of only 70 km~s$^{-1}$, found in the
  Local Group Standard of Rest. The population is in-falling at 100
  km~s$^{-1}$ toward the Local Group barycenter. These objects have a
  characteristic morphology, in which one or more compact cores is
  embedded in a diffuse halo. The compact cores typically account for
  40\% of the \hi line flux while covering some 15\% of the source
  area. The narrow line width of all core components allows unambiguous
  identification of these with the cool condensed phase of \hi$\!$, the
  CNM, with kinetic temperature near 100~K, while the halos appear to
  represent a shielding column of warm diffuse \hi$\!$, the WNM, with
  temperature near 8000~K. We detect a core with one of the narrowest
  \hi emission lines ever observed, with intrinsic FWHM of no more than
  2~km~s$^{-1}$ and 75~K brightness. From a comparison of column and
  volume densities for this feature we derive a distance in the range
  0.5 to 1 Mpc. We determine a metallicity for this same object of 0.04
  to 0.07 solar. Comparably high distances are implied by demanding the
  stability of objects with multiple cores, which show relative
  velocities as large as 70~km~s$^{-1}$ on 30~arcmin scales. Many of
  the compact cores show systematic velocity gradients along the major
  axis of their elliptical extent which are well-fit by circular
  rotation in a flattened disk system. Two out of three of the derived
  rotation curves are well-fit by Navarro, Frenk, and White (1997) cold
  dark matter profiles. These kinematic signatures imply a high
  dark-to-visible mass ratio of 10--50, for $D~=~0.7$~Mpc, which scales
  as $1/D$. The implied dark matter halos dominate the mass volume
  density within the central 2~kpc (10~arcmin) of each source,
  providing a sufficent hydrostatic pressure to allow CNM condensation.
  The CHVC properties are similar in many respects to those of the
  Local Group dwarf irregular galaxies, excepting the presence of a
  high surface brightness stellar population.

\end{abstract} 
\section{Introduction} 
\label{intro}

The possibility of an extragalactic deployment of high--velocity clouds
has been considered in various contexts by (among others) Oort
(\cite{oort66,oort70,oort81}), Verschuur (\cite{vers75}), Eichler
(\cite{eich76}), Einasto et al. (\cite{eina76}), Giovanelli
(\cite{giov81}), Bajaja et al. (\cite{baja87}), Burton (\cite{burt97}),
Wakker and van Woerden (\cite{wakk97}), Blitz et al. (\cite{blit99}),
and Braun and Burton (\cite{brau99}). Blitz et al. (\cite{blit99})
interpret several general HVC properties in terms of the hierarchical
structure formation and evolution of galaxies. In this context, the
extended HVC complexes would be the nearby objects currently undergoing
accretion onto the Galaxy, while the more compact, isolated ones would
be their distant counterparts in the Local Group environment.

It is striking that the population of anomalous velocity \hi splits
naturally into two rather distinct components when observed with the
spatial sampling and sensitivity of modern surveys, like the
Leiden/Dwingeloo Survey (LDS: Hartmann \& Burton \cite{hart97}) in the
North and the Parkes Multibeam Survey (see Putman \& Gibson
\cite{putm99}) in the South. With this new perspective, it has become
clear that the large majority of the 561 objects previously cataloged
as distinct HVCs by Wakker and van Woerden (\cite{wakk91c}) are the
local, low--contrast maxima of extended diffuse complexes with angular
sizes of tens of degrees. In addition to this diffuse, yet lumpy,
component there appear to be several hundred intrinsically
compact objects with angular sizes of about 1 degree.  Objects of
intermediate angular size appear to be rather rare.

Braun and Burton (\cite{brau99}) identified and confirmed 65 examples
of compact, isolated high--velocity clouds (the CHVCs) which plausibly
represent a homogeneous subsample of the high--velocity clouds, in a
single physical state, and arguably before their physical properties
have been strongly influenced by the radiation field of the Milky Way
or of M31, or by a gravitational encounter with one of these major
systems.  Braun and Burton showed, in particular, that the velocity
dispersion of the CHVC sample is minimized in a reference frame
consistent within the observational errors to the Local Group Standard
of Rest.  This minimization provides a quantitative demonstration of
Local--Group deployment.  Within this frame, the CHVC ensemble is
dynamically quite cold, with a dispersion of only 70~km~s$^{-1}$,
although strongly infalling into the Local Group barycenter at a
velocity of about 100~km~s$^{-1}$.

Most HVCs have been identified simply by their anomalous--velocity \hi
emission in total--power surveys at an angular resolution of 0\fdg5 or
coarser.  Information on characteristic intrinsic linear scales, on
resolved spectral properties which might reveal, for example, opacity
information, and on such kinematic properties as intrinsic widths or
possible rotation, have been largely unavailable.  Of the sample of 65
compact, isolated HVCs catalogued by Braun and Burton (\cite{brau99}),
only two had been subject to interferometric imaging.  Wakker and
Schwarz (\cite{wakk91b}) used the Westerbork array to show that both
CHVC\,114$-$10$-$430 and CHVC\,111$-$06$-$466 are characterized by a
core/halo morphology, with only about 40\% of the single--dish flux
recovered on angular scales of tens of arcmin, and, furthermore, that
the linewidths of the single--dish spectra of these two sources were
resolved into components of some 5 \kms\, width or less.  Both of the
imaged systems display systematic velocity gradients along the major
axis of an elliptical \hi distribution, which Wakker and Schwarz judged
to be suggestive of rotation in self--gravitating systems at Local
Group distances.

If the CHVC objects are in fact a population of unevolved sub-dwarf
galaxies scattered throughout the Local Group, then they might reveal
some morphological characteristics which would not be consistent with
the expectations of other suggested scenarios, in particular for
objects ejected by a galactic fountain (e.g. Shapiro and Field
\cite{shap76}, Bregman \cite{breg80}) or located within the Galactic
halo (Moore et al. \cite{moor99}).

The Westerbork imaging discussed in this paper reveals a characteristic
core/halo morphology for the CHVCs, very narrow linewidths in the
cores, and, in many cases, a signature of rotation. These properties
are strongly suggestive of self-gravitating systems at Local Group
distances, and specifically resemble the gaseous components of some 
dwarf galaxies.  On the contrary, if the CHVC objects were produced
relatively locally by an energetic mechanism responsible for a galactic
fountain, then their \hi properties might have been expected to include
large linewidths, motions not ordered by rotation, and a characteristic
morphology other than that of a core/halo.

High--resolution imaging also makes it possible to provide specific
targets for optical observations.  Deep optical imaging would help
clarify the distinction between the CHVCs and (sub--)dwarf galaxies,
and any indication of a stellar population would allow a direct
distance determination.  If the CHVCs are at Local Group distances,
then the diffuse H\,$\alpha$ emission surrounding them is expected to
be weaker than that associated with high--velocity gas in complexes
lying within the halo of the Milky Way or extending from the Magellanic
Clouds (Bland-Hawthorn \& Maloney \cite{blan99}).  Furthermore,
high--resolution imaging is necessary in order to interpret existing
observations aimed at detecting an HVC in absorption against a star in
the halo of the Milky Way or against an extragalactic background source
of continuum radiation.  In either case, the absorption experiment
targets an extremely small area.  It is not clear if the failure to
detect absorption in at least some of the published accounts should be
attributed to the HVC in question not in fact covering the continuum
source, rather than to the metallicity characterisics of the HVC (for
the cases of a negative result toward an extragalactic background
sources) or to the HVC being at a larger distance (for the cases of a
negative result toward halo stars).  The results shown below indicate
that only a small portion of each object is characterized by the high
column density \hi emission from the CHVC cores, and thus that
selection of suitable probes for absorption--line measurements must
take the actual column density distribution into account.

If the compact HVCs discussed here are distributed throughout the Local
Group as primordial objects, either surviving as remnants of the
formation of the Local Group galaxies, or still raining in to these
systems in a continuing evolution, then such objects would also be
expected near other groups of galaxies.  Several searches for such
objects beyond the Local Group are currently in progress. The results
presented here on the small angular sizes of the CHVCs generally and
the even smaller surface covering factor of the high column density
cores must be considered when designing an optimum search strategy or
when calculating the expected detection statistics for a given
experiment directed at other galaxies or groups of galaxies. A
conclusive detection experiment of this type has not yet been carried
out.

Our discussion is organized as follows.  We begin by describing the
method of sample selection in \S\ref{sec:samp}, proceed with a
description of the newly acquired observations in \S\ref{sec:data} and
continue with a presentation and discussion of our results in
\S\S\ref{sec:results} and \ref{sec:disc}. We close by briefly
summarizing our results in \S\ref{sec:summ}.

\section{Sample selection} 
\label{sec:samp}

Because high--resolution \hi imaging had previously only been obtained
for two objects of the CHVC class, we chose a sample from the catalog
of Braun and Burton (\cite{brau99}) which spanned a wide range of the
source properties, paying particular attention to source position both
on the sky and in radial velocity, as well as to the physical
attributes of linewidth and total \hi flux density. Although only six
CHVC sources were imaged in our program, the sources are distributed
widely in galactic coordinates, span radial velocities of $-275<v_{\rm
  LSR}<+165$ \kms, vary in linewidth from 6 to 95 \kms, and in line
flux from 25 to 300 Jy\,\kms. The source properties are summarized in
Table~\ref{tab:sample}.

\setlength{\tabcolsep}{4pt}
\begin{flushleft} 
\begin{table*} 
\caption[]{Compact, isolated high--velocity clouds imaged with the WSRT.} 
\label{tab:sample}
\begin{tabular}{cccccc} 
\noalign{\smallskip} \hline 
Name & RA(2000) & Dec(2000) &{\small LDS} flux & {\small LDS} 
{\small FWHM} & {\small LDS} structure \\
{\small CHVC}\,$lll \pm bb \pm vvv$ & (h m) & (\degr \, \arcmin) & 
(Jy\,\kms) & (\kms) &(a $\times$ b @ {\small PA}) \\
\noalign{\smallskip} \hline 
{\sc CHVC}\,069+04$-$223 & 19 50.1 & 33 41 & 86 & 34.0 & 1\fdg0
$\times$ 0\fdg9 @ 330$^\circ$ \\
{\sc CHVC}\,115+13$-$275 & 22 56.9 & 74 33 & 105 & 95.4 & 1\fdg1
$\times$ 0\fdg6 @ 114$^\circ$ \\
{\sc CHVC}\,125+41$-$207 & 12 24.0 & 75 36 & 235 & 5.9 & 1\fdg3
$\times$ 0\fdg7 @ 292$^\circ$ \\
{\sc CHVC}\,191+60+093 & 10 36.9 & 34 10 & 66: & 29.6 & 1\fdg0 $\times$
0\fdg9 @ 170$^\circ$ \\
{\sc CHVC}\,204+30+075 & 08 27.5 & 20 09 & 305: & 33.9 & 0\fdg8
$\times$ 0\fdg6 @ 215$^\circ$ \\
{\sc CHVC}\,230+61+165 & 10 55.2 & 15 28 & 26: & 29.3 & 1\fdg4 $\times$
0\fdg7 @ 182$^\circ$ \\
\noalign{\smallskip} \hline 
\end{tabular} 
\end{table*} 
\end{flushleft} 

\section{Data} 
\label{sec:data}

Observations of the six CHVC fields were obtained with the WSRT array
between 12/11/98 and 17/12/98. One twelve--hour integration was
obtained for each field in the standard WSRT array configuration having
a shortest baseline of 36 meters. On--source integration was bracketed
by half--hour observations of the calibration sources 3C286 and 3C48.
At the time of the observations, 11 of the 14 telescopes of the array
were equipped with the upgraded receivers having a system temperature
of about 27~K in the 1150 to 1850~MHz band. (As of 11/3/99 the entire
array is equipped with a receiver package covering the range from
250~MHz to 8~GHz with eight feed systems mounted in a prime--focus
turret.)  The interim WSRT correlator system available at the time was
used to provide 256 uniformly weighted spectral channels in two linear
polarizations across 2.5~MHz centered on the $v_{\rm LSR}$ velocity of
each source. The effective velocity resolution was 1.2 times the
channel spacing of 2.06~km~s$^{-1}$.

Standard gain and bandpass calibrations were carried out after editing
the data and deleting all baselines effected by shadowing. A component
model for the continuum emission from each field was derived from an
image made from the average of the line--free spectral channels. This
model was subtracted directly from the visibility data. The block of
spectral channels containing line emission was imaged with a
visibility--based clean deconvolution proceeding down to a flux level
of twice the rms noise level at a variety of spatial resolutions.
Uniform weighting of the visibility data was employed, together with a
series of Gaussian tapers decreasing to 30\% amplitude at projected
baselines of 1.25, 2.5, 5, and 10 k$\lambda$, resulting in spatial
resolutions of about 120, 60, 30, and 20~arcsec, respectively.  In a
few cases, some residual continuum emission was still present in the
data cubes; in those cases, several spectral channels from both edges
of the cube were averaged together and subtracted from the entire cube.

The typical rms noise levels in the deconvolved cubes were 2.0, 1.7,
1.4, and 1.3 mJy per beam per spectral channel at the four spatial
resolutions noted above. The corresponding brightness sensitivities
were, respectively, 0.085, 0.29, 0.95, and 2.0~K in a single channel of
2.06~km~s$^{-1}$ width. (Flux per beam and brightness temperature are
related as usual by $S=2k_{\rm B}T_{\rm B}\Omega_{\rm B}/\lambda^2$, or
$S_{{\rm mJy/Beam}}=0.65\Omega_{{\rm as}}T_{\rm B}/\lambda_{{\rm
    cm}}^2$ 
for $\Omega_{{\rm as}}$, the
beam area in arcsec$^2$.)  Expressed as an optically thin \hi column
density, the sensitivities correspond to 0.32, 1.1, 3.6, and
7.4~$\times 10^{18}$~cm$^{-2}$, respectively, for emission filling the
beam and extending over a single velocity channel. While some \hi
emission profiles originating in the cool atomic phase are indeed as
narrow as the single--channel velocity width (as we will see below),
emission profiles originating in the warm neutral phase will have
substantially broader linewidths. For example, the thermal linewidth of
a component with kinetic temperature, $T_{\rm k}=8000$~K, will be
21~km~s$^{-1}$ FWHM. At this velocity resolution, our column density
sensitivity is degraded to 1.0, 3.5, 11, and 24~$\times
10^{18}$~cm$^{-2}$.  This somewhat counter--intuitive correspondence of
column density sensitivity with the expected linewidth must be borne in
mind. Narrow lines are significantly easier to detect than broad ones.

Moment images (zero, first, and second) were generated from each cube
after employing a blanking criterion for inclusion of each pixel in the
weighted sum. This involved demanding a brightness in excess of about
2$\sigma$ after smoothing the cube by a factor of three both spatially
and in velocity. Images of integrated emission were corrected for the
primary--beam response of the instrument, which is well--approximated
(at 1420 MHz) by a circular Gaussian with 2110~arcsec FWHM.

While structures extending over as much as 10~arcmin in a single
spectral channel were adequately recovered in the resulting images,
there were also indications of the presence of more diffuse features
which could not be adequately imaged. These were apparent due to the
artifacts they induced; specifically, the so--called short--spacing
bowl surrounding regions of bright extended emission. Even with careful
windowing during the deconvolution it was not possible to completely
eliminate such artifacts. Consequently, a significant fraction of the
total flux detected in the single--dish observations could not be
recovered.  The integrated \hi flux detected in the reconstructed
images after primary beam correction varied from less than 1\% to as
much as 55\% of that detected in the LDS. The percentages of total
fluxes contributed by the dense cores are indicated for each
observed field in Table~\ref{tab:properties}.

\setlength{\tabcolsep}{4pt}
\begin{flushleft} 
\begin{table*} 
\caption[]{Properties of CHVCs imaged with the WSRT.} 
\label{tab:properties}
\begin{tabular}{ccccc} 
\noalign{\smallskip} \hline 
Name & {\small WSRT} flux & Core
flux & Core area & Distance \\
{\small CHVC}\,$lll \pm bb \pm vvv$ & 
(Jy\,\kms) & \% & \% & (kpc)\\
\noalign{\smallskip} \hline 
{\sc CHVC}\,069+04$-$223 &  21.9 & 26 & 3.4 & $>$(40--50)\\
{\sc CHVC}\,115+13$-$275 & 20.7 & 20 & 10. & $>$(40-100)\\
{\sc CHVC}\,125+41$-$207 & 130. & 55 & 25 & 500--1000\\
{\sc CHVC}\,191+60+093 & 0.22 & 0.3 & 0.2 & $>$7: \\
{\sc CHVC}\,204+30+075 & 116. & 38 & 48 & $>$(40--130)\\
{\sc CHVC}\,230+61+165 & 0.65 & 2.5 & 0.3 & $>$4:\\
\noalign{\smallskip} \hline 
\end{tabular} 
\end{table*} 
\end{flushleft} 

A crude attempt was made to correct the interferometric images of
integrated \hi for the non--detection of diffuse emission features. An
elliptical Gaussian with dimensions and orientation as measured in the
LDS and an integral flux sufficient to recover the LDS total was added
to the interferometric images. These parameters are summarized for the
six fields in Table~\ref{tab:properties}. The resulting composite images
are presented in turn below. The lowest column--density contours in
these composite images follow the elliptical outlines of the LDS source
model, while the higher levels are dominated by the compact structures
detected interferometrically. Although not appropriate for detailed
analysis on spatial scales between about 10 and 30 arcmin, these
composite images are consistent with all of the data currently in hand
and are indicative of the column densities likely to be present in
diffuse structures as well as their overall spatial extent. In at
least one case (discussed below) we were able to verify with
independent data that the resulting composite provided a reasonable
representation of the source structure and extent.

\section{Results}
\label{sec:results}

In this section, we briefly summarize the structural and kinematic data 
obtained for each of the six observed fields using the WSRT.

\subsection{CHVC\,069+04$-$223}

\begin{figure*} 
\resizebox{18cm}{!}{\includegraphics{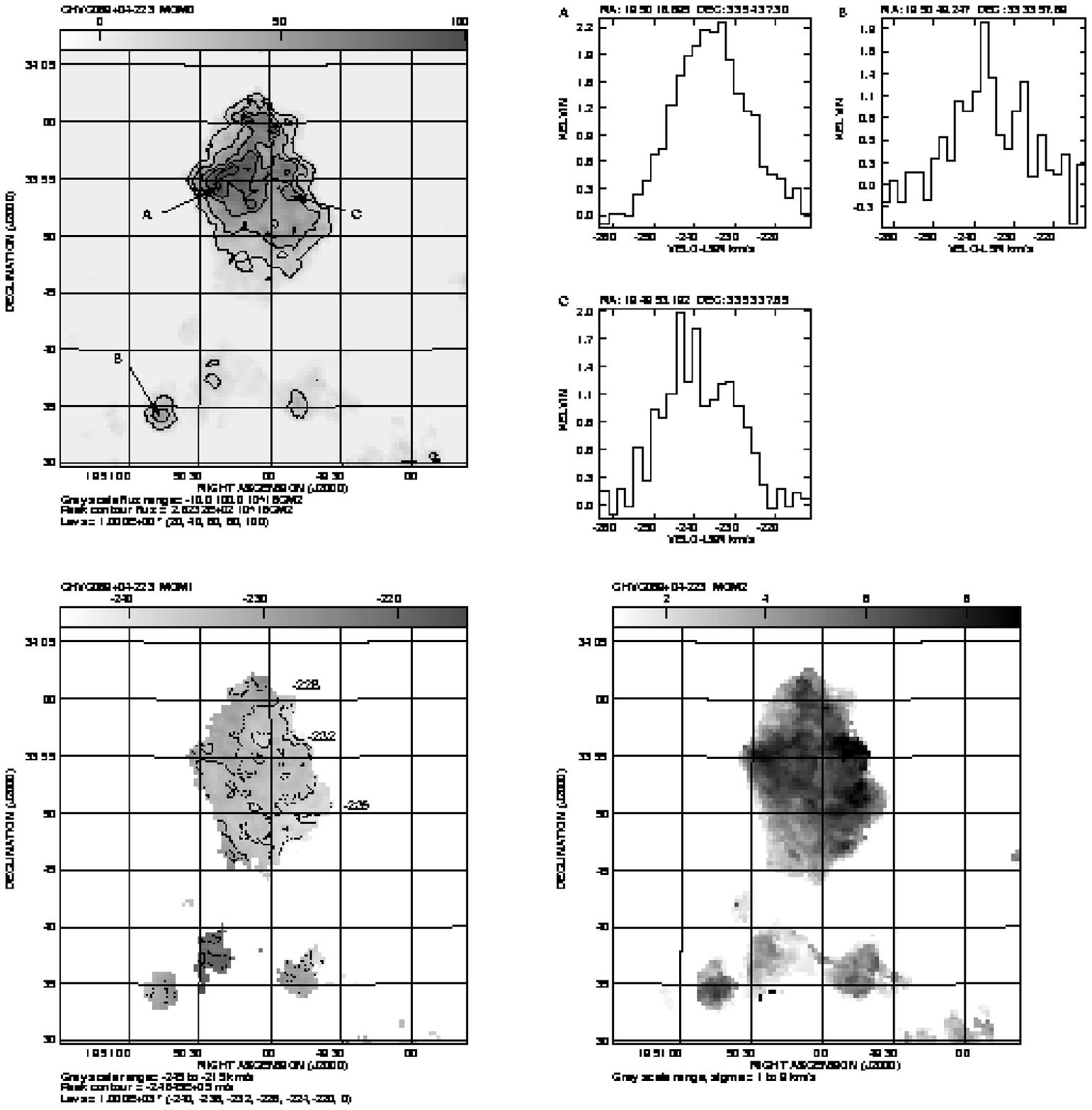}}
\caption{ Imaging data for CHVC\,069+04$-$223 at 1~arcmin and 2~km~s$^{-1}$
  resolution. Upper left panel: apparent integrated \hi (assuming
  negligible opacity), with contours at 20, 40, 60, 80, and 100
  $\times10^{18}$ cm$^{-2}$ and a linear grey--scale extending from $-10$
  to 100$\times10^{18}$ cm$^{-2}$. Upper right panel: brightness
  temperature spectra at the indicated positions.   Lower left panel:
  intensity weighted line--of--sight velocity, $v_{\rm LSR}$, with contours
  at $-240, -236, -232, -228, -224,$ and $-$220~km~s$^{-1}$ and a linear
  grey--scale extending from $-245$ to $-215$~km~s$^{-1}$. Lower right
  panel: intensity weighted distribution of squared velocity,
  corresponding to the velocity dispersion of a Gaussian profile, with
  a linear grey--scale extending from 1 to 9~km~s$^{-1}$.
 } \label{fig:h069o}
\end{figure*} 

\begin{figure*}
\resizebox{12cm}{!}{\includegraphics{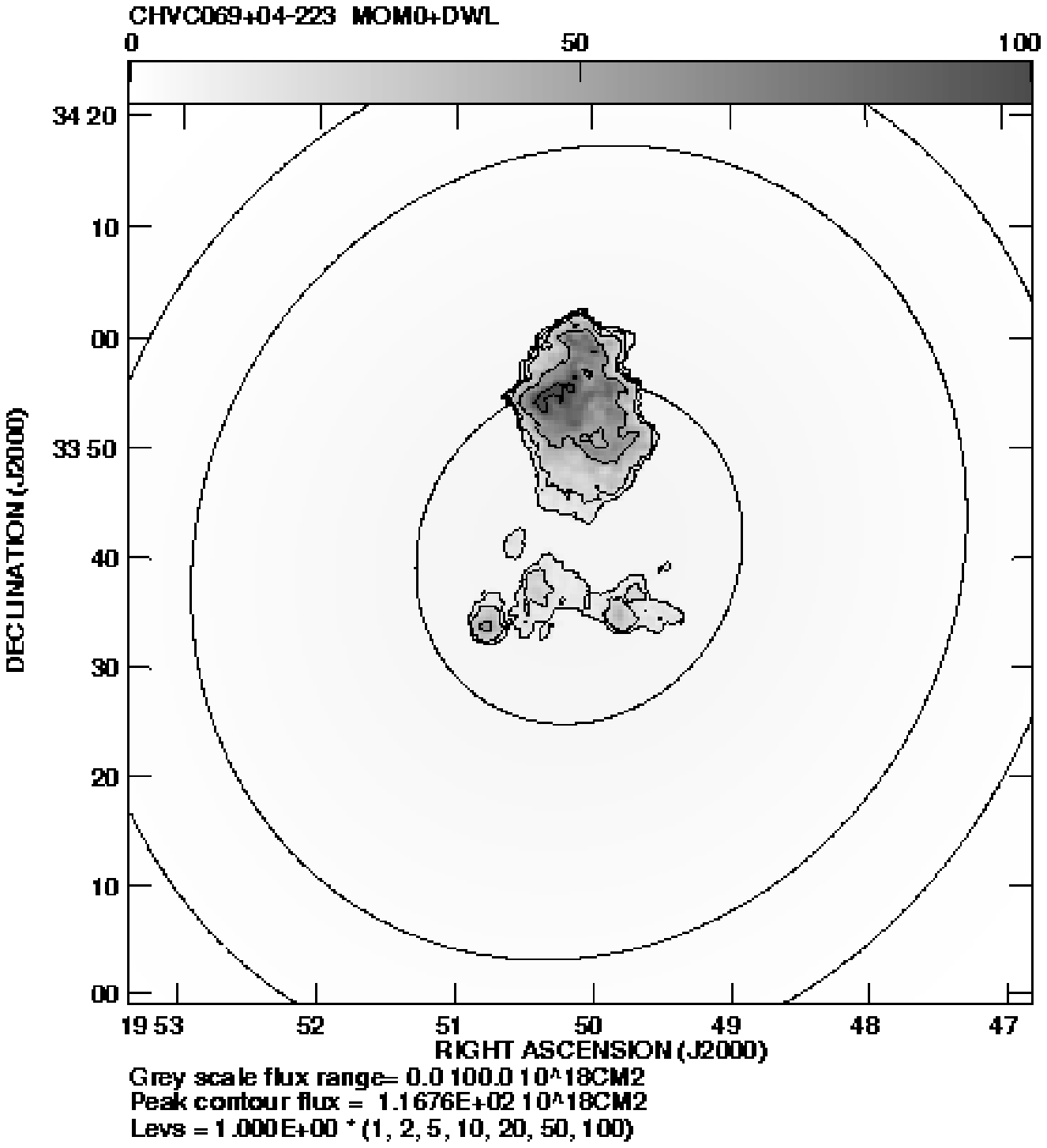}}
\hfill
\parbox[b]{55mm}{
\caption{Column density distribution of \hi in
  CHVC\,069+04$-$223 at 1~arcmin resolution reconstructed from LDS and WSRT
  data. Contours are drawn at 1, 2, 5, 10, 20, 50, and 100
  $\times10^{18}$ cm$^{-2}$;  a linear grey--scale extends from $0$
  to 100$\times10^{18}$ cm$^{-2}$. 
 } \label{fig:h069m}}
\end{figure*}

The moment images and a few representative spectra for this field are
shown at 1~arcmin resolution in Fig.~\ref{fig:h069o}.  The integrated
\hi distribution is dominated by a bright elongated concentration
(clump~A) of some 15~arcmin extent. Several fainter knots lie about
10~arcmin further to the South, and are connected to each other by a
low brightness bridge of emission (more clearly seen in
Fig.~\ref{fig:h069m}). Spectra toward some of the brightest emission
regions all have ($i$) peak brightnesses of about 2~K, ($ii$)
relatively broad line profiles of about 15~km~s$^{-1}$ FWHM, and
($iii$) a centroid velocity near $-$235~km~s$^{-1}$.  At several local
$N_{\rm HI}$ maxima within clump~A the profiles become doubly peaked.
These will be referred to again below.  The velocity field displays at
least one systematic trend: a moderate velocity gradient, oriented
along PA 30$^\circ$ (East of North) along the major axis of clump~A
from about $-$230 to $-$240~km~s$^{-1}$ and extending over some
10~arcmin.

The distribution of profile linewidth shows that clump~A has a
moderately high linewidth throughout, but also has several distinct
regions with apparent dispersions as high as 8~km~s$^{-1}$. These
correspond to regions of line splitting rather than to simple line
broadening, while the overall velocity centroid is unchanged. Two such
regions are observed, centered near $(\alpha,\delta)=(19^{\rm h}49^{\rm
  m}50^{\rm s},\,33^\circ52^\prime)$ and $(\alpha,\delta)=(19^{\rm
  h}49^{\rm m}55^{\rm s},\,33^\circ54^\prime)$.  The two peaks in the
line profile are separated by about 10~km~s$^{-1}$ in both cases, while
the angular extent is only some 90~arcsec.

The integrated \hi distribution in CHVC\,069+04$-$223, after inclusion of
a simple representation of the diffuse emisison derived from the LDS
data, is shown in Fig.~\ref{fig:h069m}.  The brightest knots have
column densities just in excess of 10$^{20}$~cm$^{-2}$, while the
diffuse underlying envelope peaks at about $5\times10^{18}$ cm$^{-2}$.

\subsection{CHVC\,115+13$-$275}

\begin{figure*}
\resizebox{18cm}{!}{\includegraphics{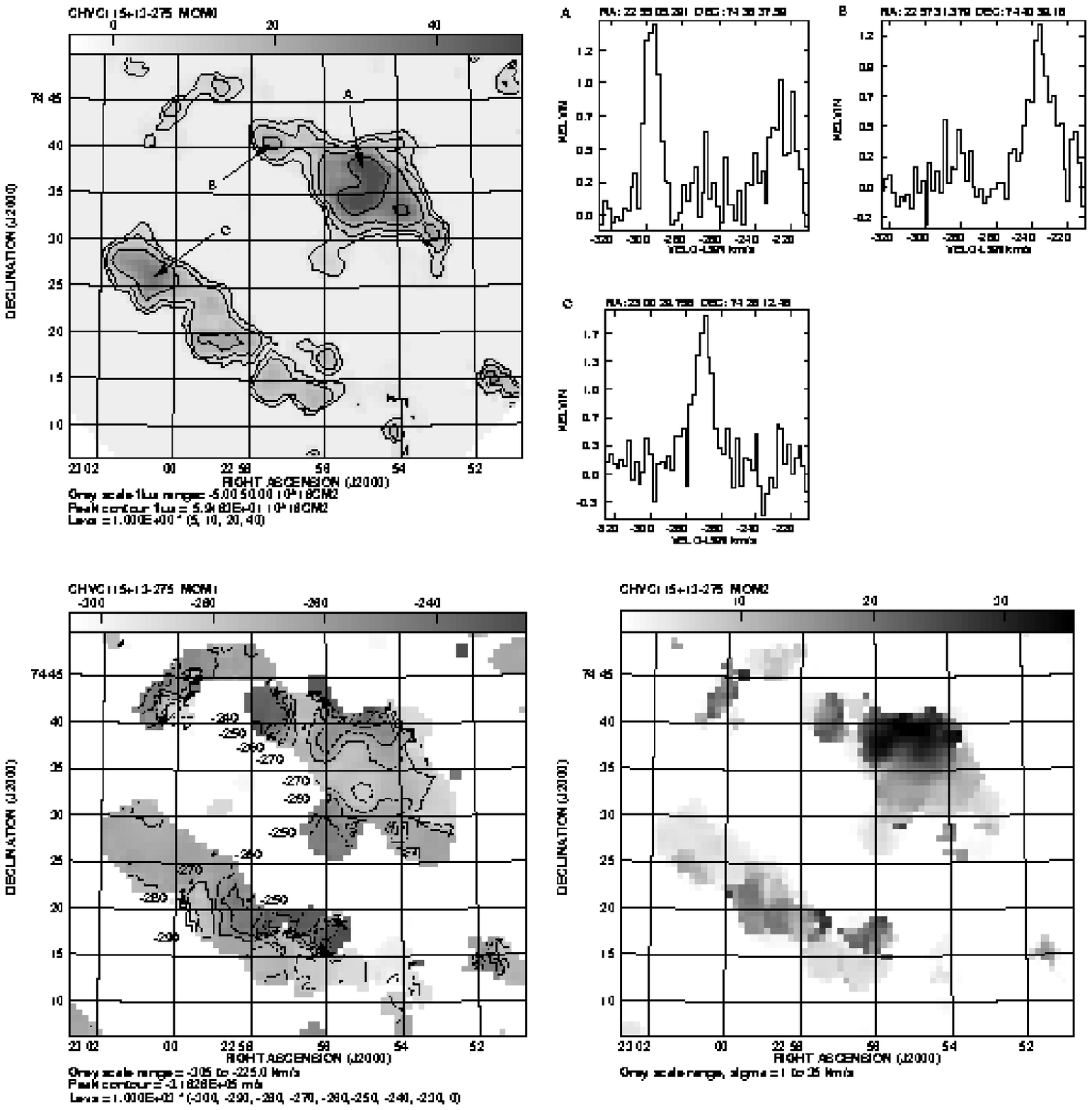}}
\caption{ Imaging data for CHVC\,115+13$-$275 at 2~arcmin and 2~km~s$^{-1}$
  resolution. Upper left panel: apparent integrated \hi (assuming
  negligible opacity), with contours at 5, 10, 20, and 40
  $\times10^{18} $cm$^{-2}$ and a linear grey--scale extending from $-5 $
  to 50$\times10^{18} $cm$^{-2}$. Upper right panel: brightness
  temperature spectra at the indicated positions. Lower left panel:
  intensity weighted line--of--sight veloctiy, $v_{\rm LSR}$, with contours
  at $-$300 to $-$230~km~s$^{-1}$ in steps of 10~km~s$^{-1}$ and a linear
  grey--scale extending from $-305$ to $-225$~km~s$^{-1}$.  Lower right
  panel: intensity weighted distribution of squared velocity,
  corresponding to the velocity dispersion of a Gaussian profile, with
  a linear grey--scale extending from 1 to 35~km~s$^{-1}$.
 } \label{fig:h115o}
\end{figure*} 

The basic data for this field are shown at 2~arcmin resolution in
Fig.~\ref{fig:h115o}.  This object is resolved in the WSRT data into a
collection of approximately 10 sub--structures. Each clump is between
about 1 and 10~arcmin in size, while the ensemble of clumps is
distributed over a region of about 30~arcmin diameter. Spectra towards
several of the clumps show that each feature has a peak brightness of
about 1~K and an intrinsic velocity width of between 10 and
15~km~s$^{-1}$ FWHM.  On the other hand, each feature has a distinct
centroid velocity so that the collection spans the interval of $-$300
to $-$220~km~s$^{-1}$. The global distribution of the line--of--sight
velocities shows no obvious pattern. High, low, and intermediate
velocity clumps are distributed across the field. Some of the most
extreme velocities do occur in close spatial proximity, however, near
$(\alpha,\delta)=(23^{\rm h}55^{\rm m},\,74^\circ38^\prime)$; but
comparable velocities are also seen elsewhere.

More careful examination of the individual clumps reveals that many of
them have significant velocity gradients oriented preferentially along
their long axes. This is true for the clumps centered near
$(\alpha,\delta)=(22^{\rm h}55^{\rm m},\,74^\circ32^\prime)$,
$(23^{\rm h}01^{\rm m},\,74^\circ27^\prime)$, $(22^{\rm h}57\fm5,\,
74^\circ42^\prime)$, and
$(22^{\rm h}57\fm5,\,74^\circ15^\prime)$. The typical gradient observed 
is some 10~km~s$^{-1}$ over an angular extent of 5~arcmin.

The image of velocity dispersion shown in the lower--left panel of Fig. 3 
serves primarily to illustrate the
regions where multiple velocity components overlap along the
line of sight. The morphology of the regions having multiple--velocity
components does not seem peculiar, but instead suggests simply
line--of--sight overlap of distinct features rather than a physical
interaction.

\begin{figure*}
\resizebox{12cm}{!}{\includegraphics{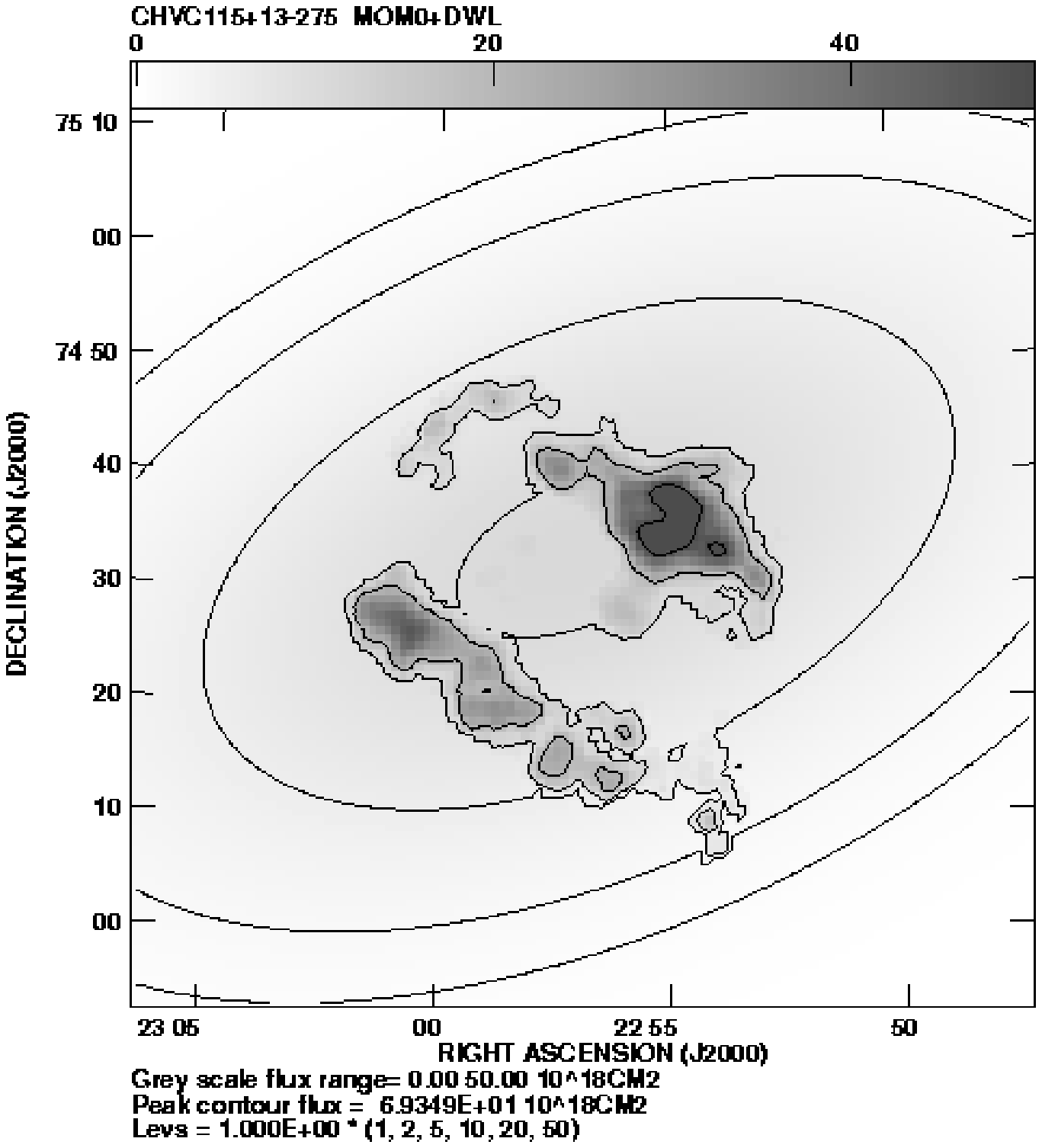}}
\hfill
\parbox[b]{55mm}{
\caption{ Column density distribution of \hi in
  CHVC\,115+13$-$275 at 2~arcmin resolution reconstructed using LDS and WSRT
  data. Contours are drawn at 1, 2, 5, 10, 20, and 50 
  $\times10^{18}$~cm$^{-2}$ and a linear grey--scale extends from $0$
  to 50$\times10^{18}$~cm$^{-2}$.
 } \label{fig:h115m}}
\end{figure*} 

The composite image of integrated \hi in CHVC\,115+13$-$275 is shown in
Fig.~\ref{fig:h115m}. Column densities in the knots do not exceed
about $5\times10^{19}$~cm$^{-2}$, while the broad underlying 
distribution peaks between 5 and $10\times10^{18}$~cm$^{-2}$

\subsection{CHVC\,125+41$-$207}

\begin{figure*}
\resizebox{18cm}{!}{\includegraphics{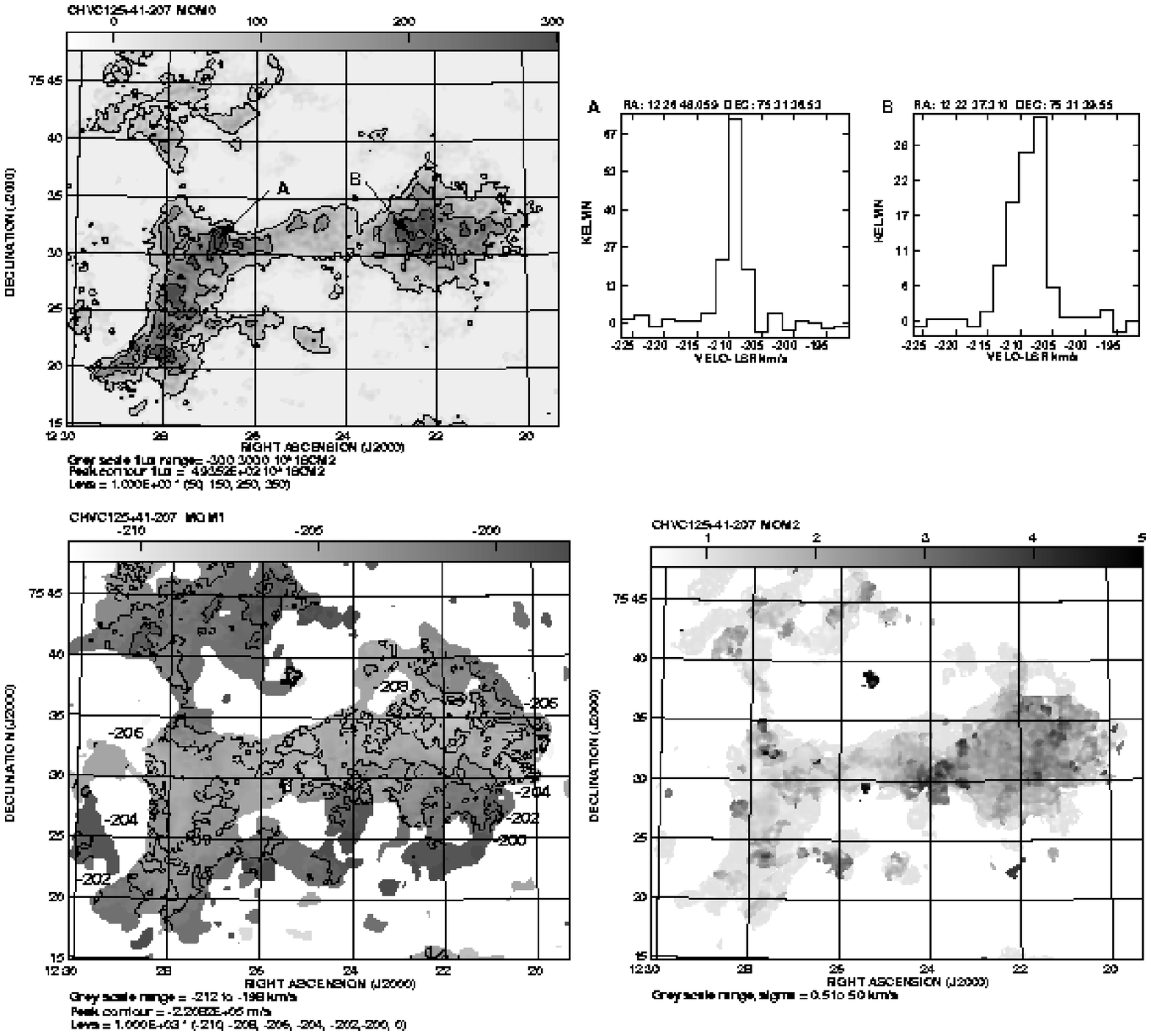}}
\caption{  Imaging data for CHVC\,125+41$-$207 at 28~arcsec and 2~km~s$^{-1}$
  resolution. Upper left panel: apparent integrated \hi (assuming
  negligible opacity), with contours at 50, 150, 250, and 350
  $\times10^{18}~$cm$^{-2}$ and a linear grey--scale extending from $-30$
  to 300$\times10^{18}~$cm$^{-2}$. Upper right panel: brightness
  temperature spectra at the indicated positions. The extremely narrow
  linewidths --- the spectrum shown on the left in this panel is one of the
  narrowest \hi emission lines ever observed --- robustly constrain the 
  kinetic temperatures. Lower left panel:
  intensity weighted line-of-sight veloctiy, $v_{\rm LSR}$, with contours
  at $-210, -208, -206, -204, -202$, and $-$200~km~s$^{-1}$ and a linear
  grey--scale extending from $-212$ to $-198$~km~s$^{-1}$. Lower right
  panel: intensity weighted distribution of squared velocity,
  corresponding to the velocity dispersion of a Gaussian profile, with
  a linear grey--scale extending from 0.5 to 5~km~s$^{-1}$.
 } \label{fig:h125o}
\end{figure*} 

The compact high--velocity cloud CHVC\,125+41$-$207 is particularly 
interesting.  Because of the high brightness of this source, higher angular
resolution could be employed for the data presentation.  Basic data at
28~arcsec resolution are shown in Fig.~\ref{fig:h125o}. The source
has a complex filamentary morphology extending over some 45~arcmin.
Several compact cores of only 1 to 2~arcmin extent are seen within the
Eastern segment (clump~A) of the complex. The spectrum toward the
brightest compact core is quite remarkable, in having a peak
brightness of 75~K as well as a linewidth which is completely
unresolved with our effective velocity resolution of 2.47~km~s$^{-1}$.
This width is one of the narrowest ever measured for the \hi emission
line. Velocity channels adjacent to the line peak have intensities down to
about 20\% of maximum while being spaced by only 2.06~km~s$^{-1}$. A
spectrum toward the more diffuse western knot (clump~B) has a more
modest peak brightness of some 30~K and a linewidth of about
6~km~s$^{-1}$ FWHM. The most prominent systematic trend in the velocity
field is a North--South gradient over clump~B of about 8~km~s$^{-1}$
over 10~arcmin. Clump~A has a shallower North--South velocity gradient
of about half this magnitude.
The fainter filaments have velocities which do not
follow this simple pattern. 

\begin{figure*}
\resizebox{12cm}{!}{\includegraphics{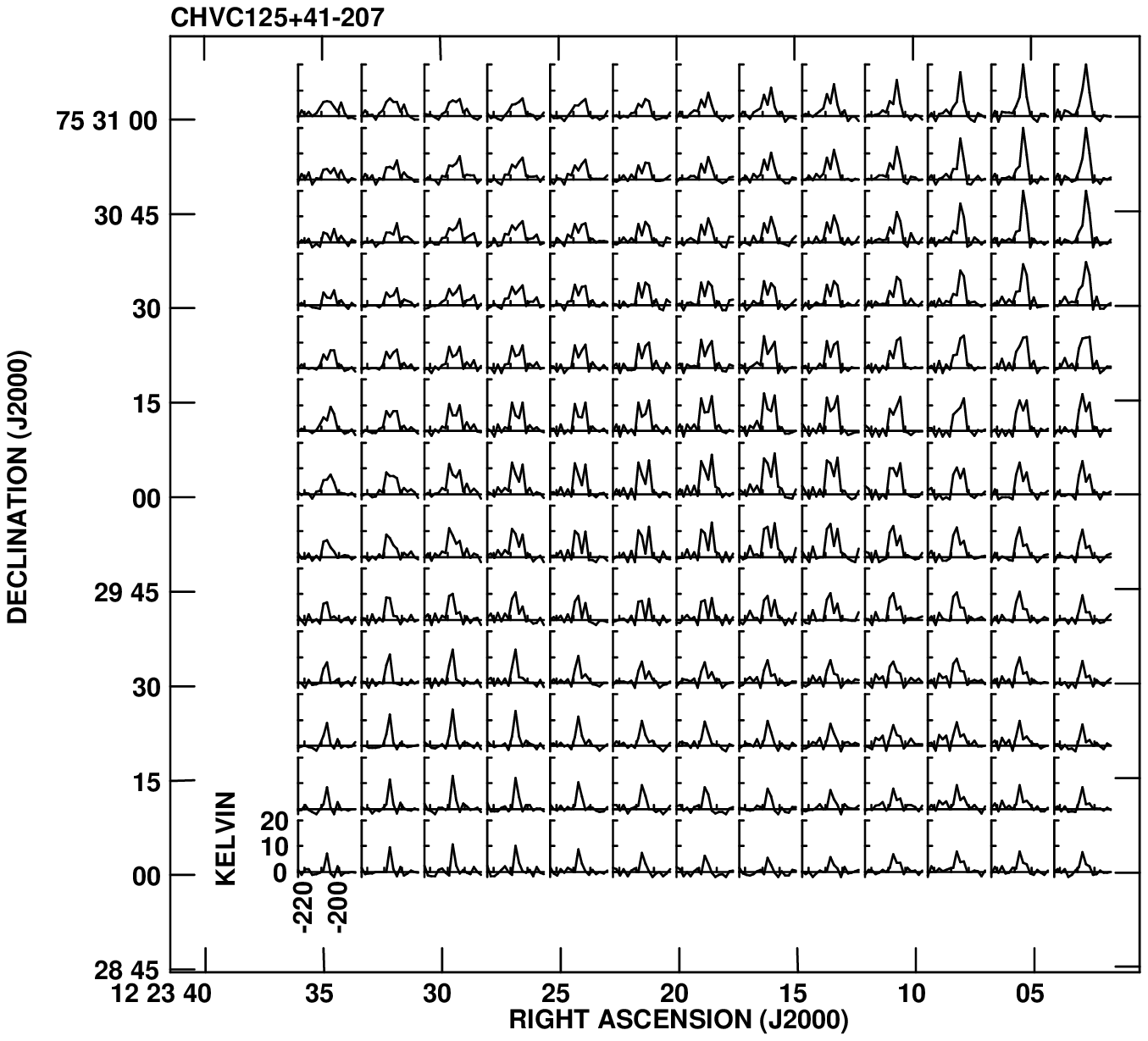}}
\hfill
\parbox[b]{55mm}{
\caption{ Brightness temperature spectra for a small region 
  (2$\times$2~arcmin) of CHVC\,125+41$-$207 at 28~arcsec and
  2~km~s$^{-1}$ resolution. Note the localized region of line splitting
  extending over about 1~arcmin.  }
\label{fig:h125s}}
\end{figure*}

The image of velocity dispersion delineates those regions with
extremely narrow linewidth from those which are only moderately narrow.
Both of the compact cores in the Eastern complex as well as several
other more diffuse structures are completely unresolved in velocity.
Clump~B is less extreme in this regard. One small region of clump~B,
centered near $(\alpha,\delta)=(12^{\rm h}23^{\rm m}18^{\rm s},\,
75^\circ30^\prime)$, deserves special mention. This minor local maximum
in the $N_{\rm HI}$ distribution displays clear velocity splitting in
the emission profile, amounting to some 5~km~s$^{-1}$ over a region of
about an arcmin in diameter, as shown in Fig.~\ref{fig:h125s}. A second
region displaying a similar phenomenon is centered near
$(\alpha,\delta)= (12^{\rm h}21^{\rm m}35^{\rm
  s},\,75^\circ32^\prime)$. Both regions are about 90~arcsec in
diameter.

The composite distribution of {\it apparent} \hi column density is
shown in Fig.~\ref{fig:h125m}. The emphasis on the term
``apparent'' is particularly apt: as we will
see below, the compact cores are likely to have opacities of about 2,
so that actual column densities are about twice the apparent ones in
these directions. Even so, the apparent column density exceeds
$4\times10^{20}$~cm$^{-2}$ over much of the source, while the
underlying halo reaches values of more than $5\times10^{18}$~cm$^{-2}$
near the knots.

\begin{figure*}
\resizebox{16cm}{!}{\includegraphics{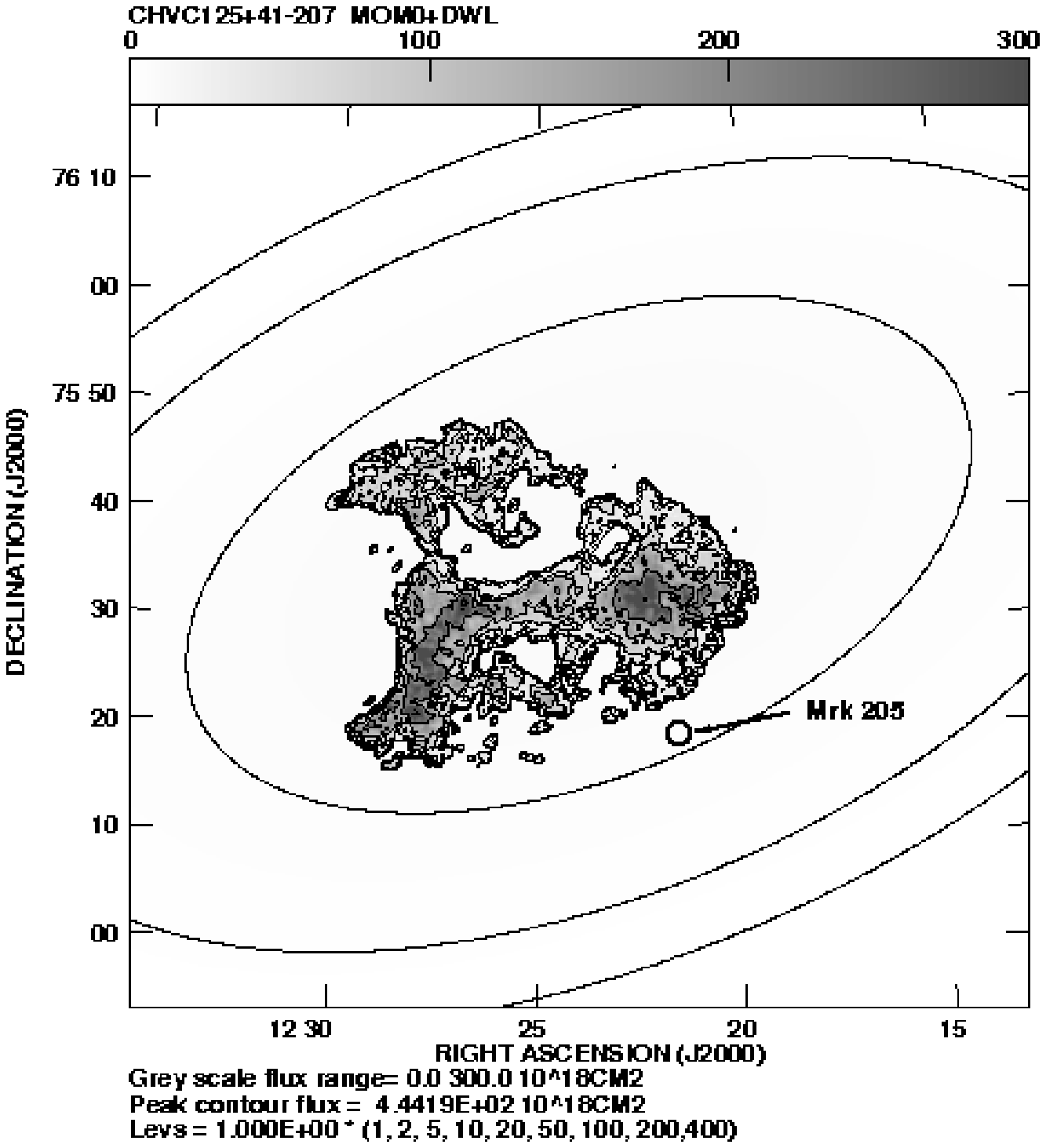}}
\caption{ Column density distribution of \hi in
  CHVC\,125+41$-$207 at 28~arcsec resolution reconstructed from the 
  LDS and WSRT
  data. Contours are drawn at 1, 2, 5, 10, 20, 50, 100, 200, and 400 
  $\times10^{18}$ cm$^{-2}$; a linear grey--scale extends from $0$
  to 300$\times10^{18}$ cm$^{-2}$.
 } \label{fig:h125m}
\end{figure*}

In this particular case, we can obtain some quality assessment of our
simple composite image of $N_{\rm HI}$, in which we have attempted to
compensate for the limited sensitivity to diffuse structures inherent in our
interferometric data.  This source has recently been observed with the
Lovell Telescope at Jodrell Bank by de~Vries et al.  (\cite{devr99}).
Comparison of the $5\times10^{18}$~cm$^{-2}$ contour in the Jodrell
Bank image with our composite image shows good agreement in both source
size and orientation.

\subsection{CHVC\,204+30+075}

\begin{figure*}
\resizebox{18cm}{!}{\includegraphics{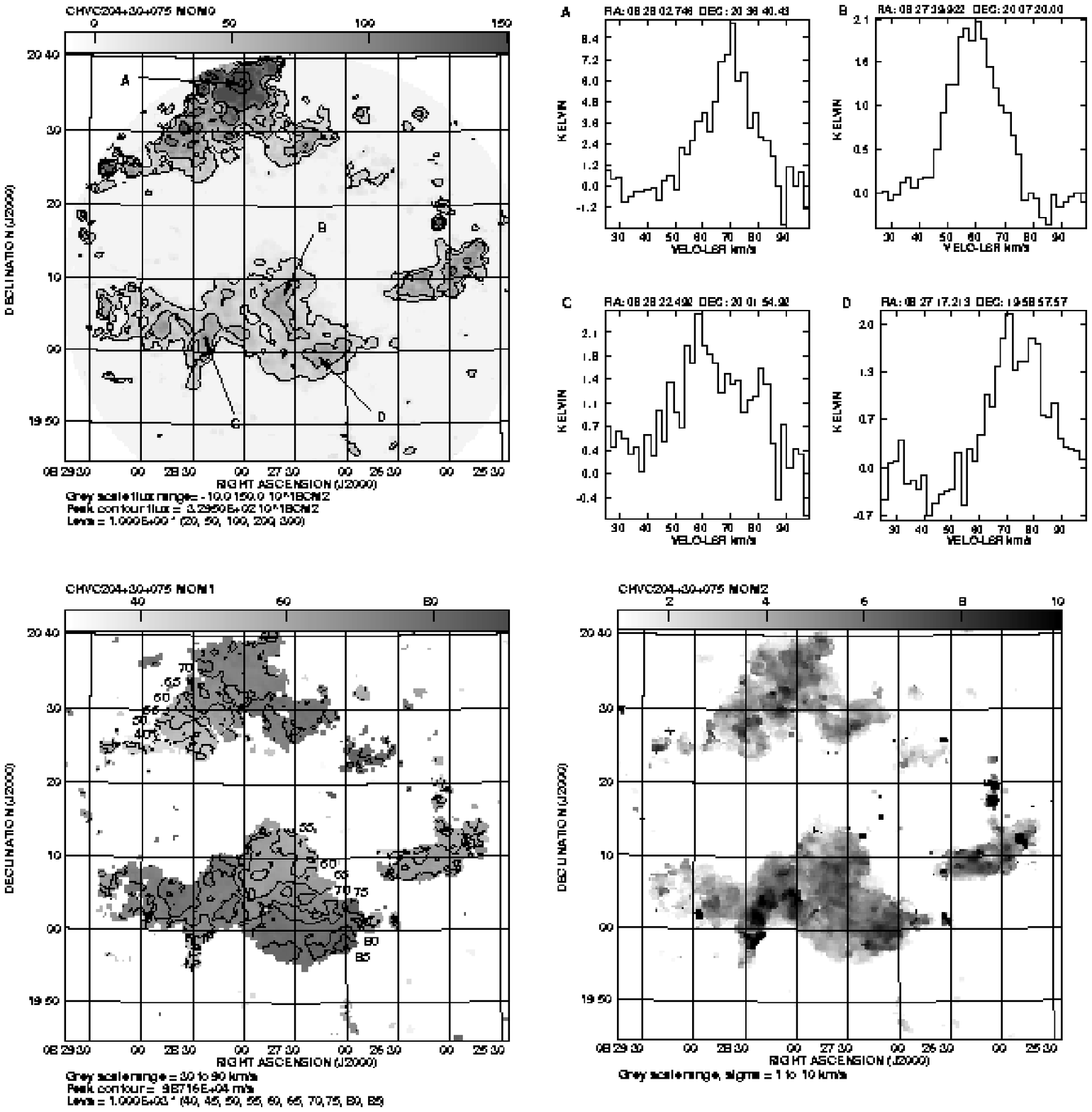}}
\caption{ Imaging data for CHVC\,204+30+075 at 1~arcmin and 2~km~s$^{-1}$
  resolution. Upper left panel: apparent integrated \hi (assuming
  negligible opacity), with contours at 20, 50, 100, 200, and 300
  $\times10^{18}$ cm$^{-2}$ and a linear grey--scale extending from $-10$
  to 150$\times10^{18}$ cm$^{-2}$. Upper right panel: brightness
  temperature spectra at the indicated positions. Lower left panel:
  intensity weighted line--of--sight velocity, $v_{\rm LSR}$, with contours
  at 40 to 85~km~s$^{-1}$ in steps of 5~km~s$^{-1}$ and a linear
  grey--scale extending from 30 to 90~km~s$^{-1}$. Lower right
  panel: intensity weighted distribution of squared velocity,
  corresponding to the velocity dispersion of a Gaussian profile, with
  a linear grey--scale extending from 1 to 10~km~s$^{-1}$.
 } \label{fig:h204o}
\end{figure*}

The data for this compact HVC are shown at 1~arcmin resolution in
Fig.~\ref{fig:h204o}. Several moderately bright clumps are seen over a
region of some 35~arcmin in size. The large angular extent of this
object in a given velocity channel challenges reconstruction of the
brightness distribution with the single 12--hour WSRT coverage which we
obtained. Some questions remain regarding the reconstruction fidelity,
particularly in regard to the most Northerly feature which extends down
to the 10\% level of the primary beam, and as such, will have had any
imperfections magnified by almost a factor of 10 in the displayed
result.  The spectra toward the more compact local maxima have peak
brightnesses between 2~K and 8~K and linewidths which are generally
some 15~km~s$^{-1}$ FWHM. Doubly--peaked spectra are seen in two
regions; in one case where two components seem to overlap along the
line of sight and in another where localized line splitting is observed
(as noted below). The large elliptical feature in the South--central
part of the field (clump~A) shows a well--defined North--South velocity
gradient running from about 55 to 80~km~s$^{-1}$ over some 12~arcmin.
The large elliptical feature in the North--East (clump~B) has a similar
gradient along PA $-$35$^\circ$ running from 45 to 70~km~s$^{-1}$ over
20~arcmin. Four other major clumps with more nearly discrete
line--of--sight velocities fill out this system.  Velocity dispersions
are typically substantially less than 10 \kms, except near
$(\alpha,\delta)=(08^{\rm h}28^{\rm m}30^{\rm s},\, 20^\circ)$ where
clump~A appears to overlap with the smaller South--Easterly clump, and
near $(\alpha,\delta)=(08^{\rm h}26^{\rm m}06^{\rm s},\,
20^\circ17^\prime30^{\prime\prime})$ where two velocity peaks separated
by about 20~km~s$^{-1}$ are seen over the 60~arcsec extent of this
local maxima in $N_{\rm HI}$.

\begin{figure*}
\resizebox{12cm}{!}{\includegraphics{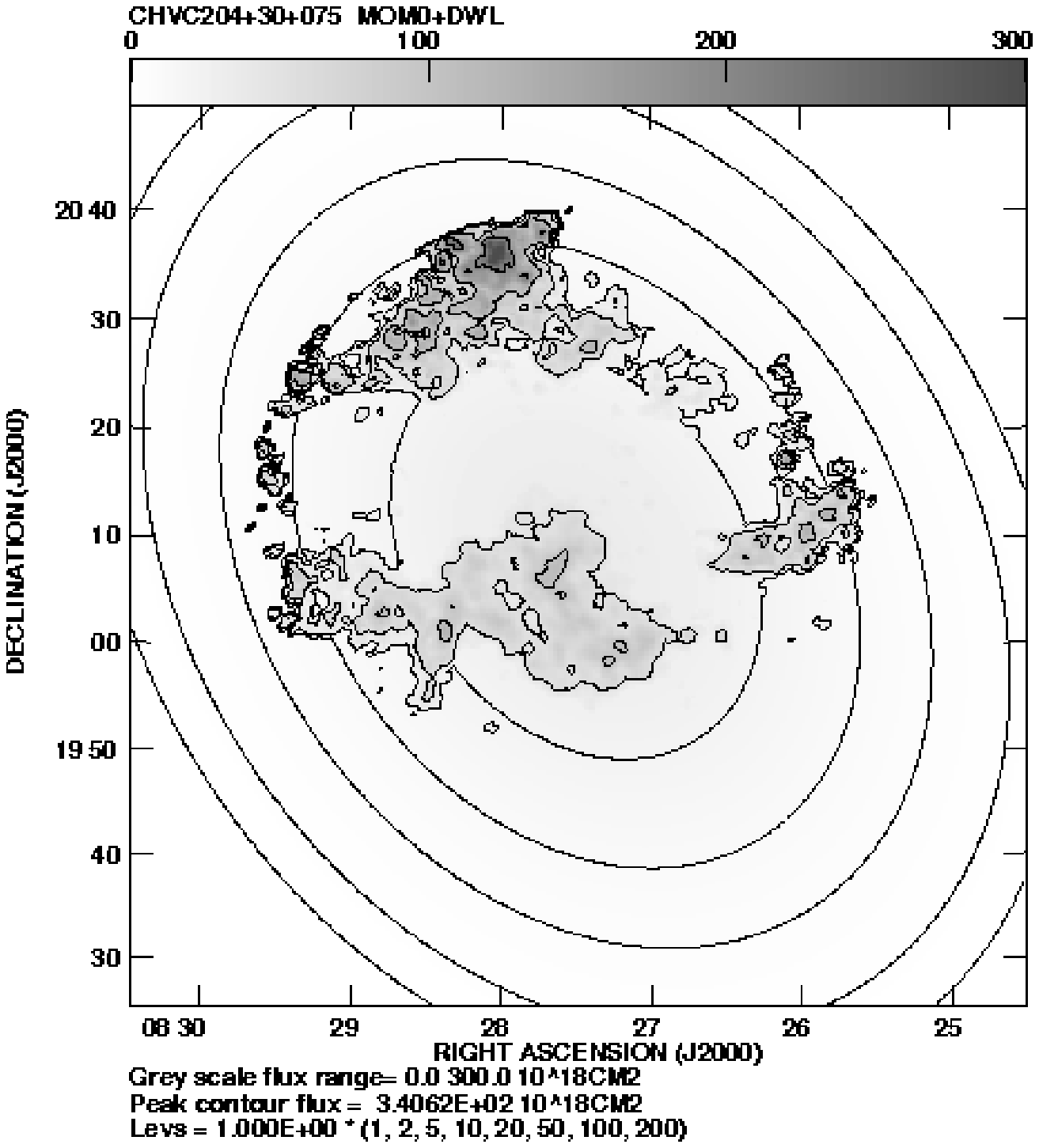}}
\hfill
\parbox[b]{55mm}{
\caption{ Column density distribution of \hi in
  CHVC\,204+30+075 at 1~arcmin resolution reconstructed from LDS and WSRT
  data. Contours are drawn at 1, 2, 5, 10, 20, 50, 100, and 200 
  $\times10^{18}$ cm$^{-2}$; a linear grey--scale extends from $0$
  to 300$\times10^{18}$ cm$^{-2}$.
 } \label{fig:h204m}}
\end{figure*} 

Column densities of a few times $10^{20}$~cm$^{-2}$ are seen
toward the brighter knots in the composite reconstruction shown in
Fig. \ref{fig:h204m}. A very substantial diffuse component is inferred
for this field, reaching column densities of a few times
$10^{19}$~cm$^{-2}$ over about 30~arcmin near the source centroid.

\subsection{CHVC\,191+60+093 and CHVC\,230+61+165}

\begin{figure*}
\resizebox{13.5cm}{!}{\includegraphics{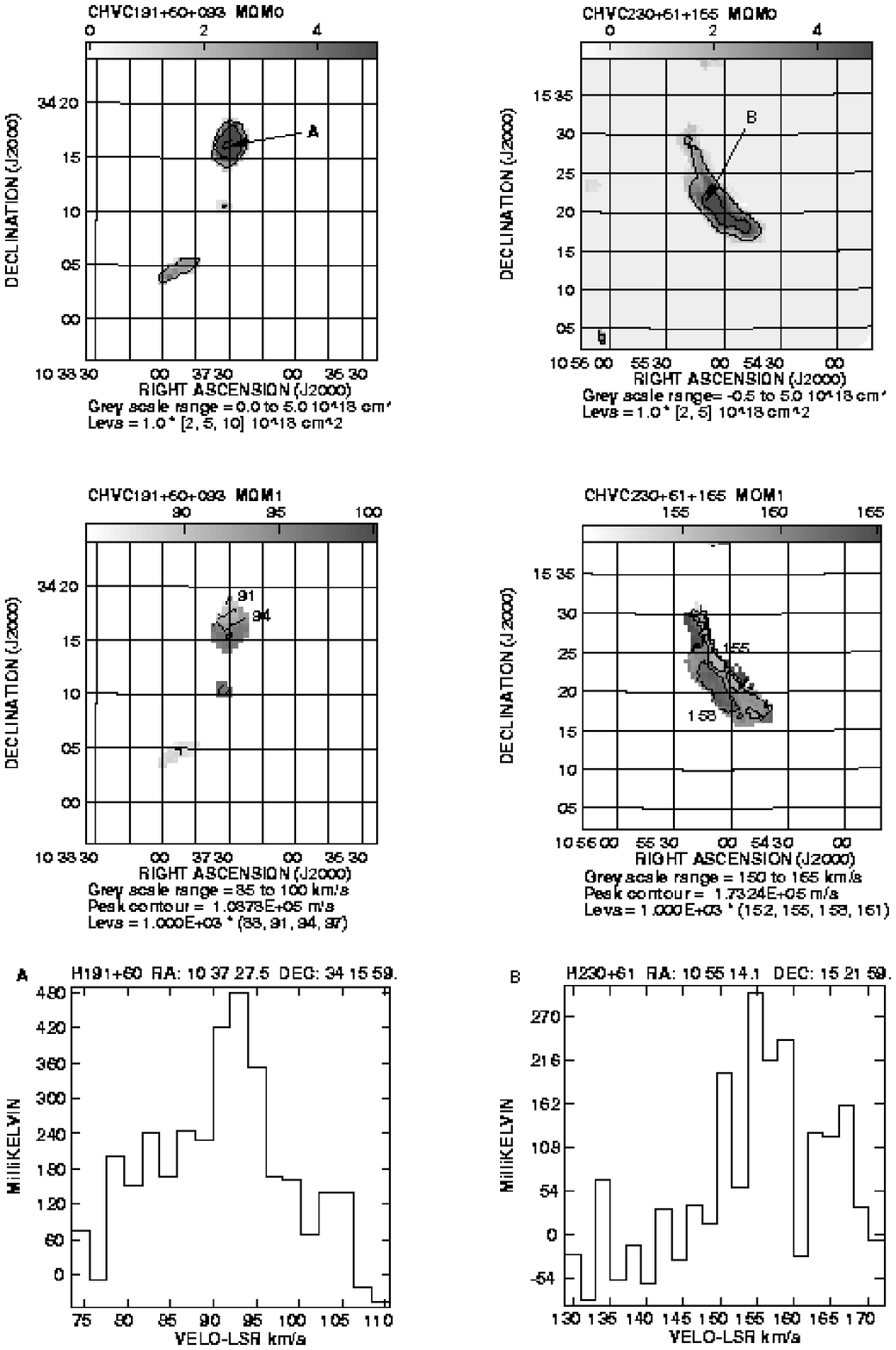}}
\caption{ Imaging data for CHVC\,191+60+093 and CHVC\,230+61+165 at
  2~arcmin and 2~km~s$^{-1}$ resolution. Upper panels: apparent
  integrated \hi (assuming negligible opacity), with contours at 2, 5,
  and 10 $\times10^{18}$ cm$^{-2}$ and a linear grey--scale extending
  from $0$ to 5$\times10^{18}$ cm$^{-2}$. Middle panels: intensity
  weighted line--of--sight velocities, $v_{\rm LSR}$, with contours at 88, 91,
  94, and 97~km~s$^{-1}$ and a linear grey--scale extending from 85 to
  100~km~s$^{-1}$ for CHVC\,191+60+093, and with contours at 152, 155, 158, and
  161~km~s$^{-1}$ and a linear grey--scale extending from 150 to
  165~km~s$^{-1}$ for CHVC\,230+61+165. Lower panels: brightness
  temperature spectra at the indicated positions.
  } \label{fig:h191o}
\end{figure*} 

\begin{figure*}
\resizebox{12cm}{!}{\includegraphics{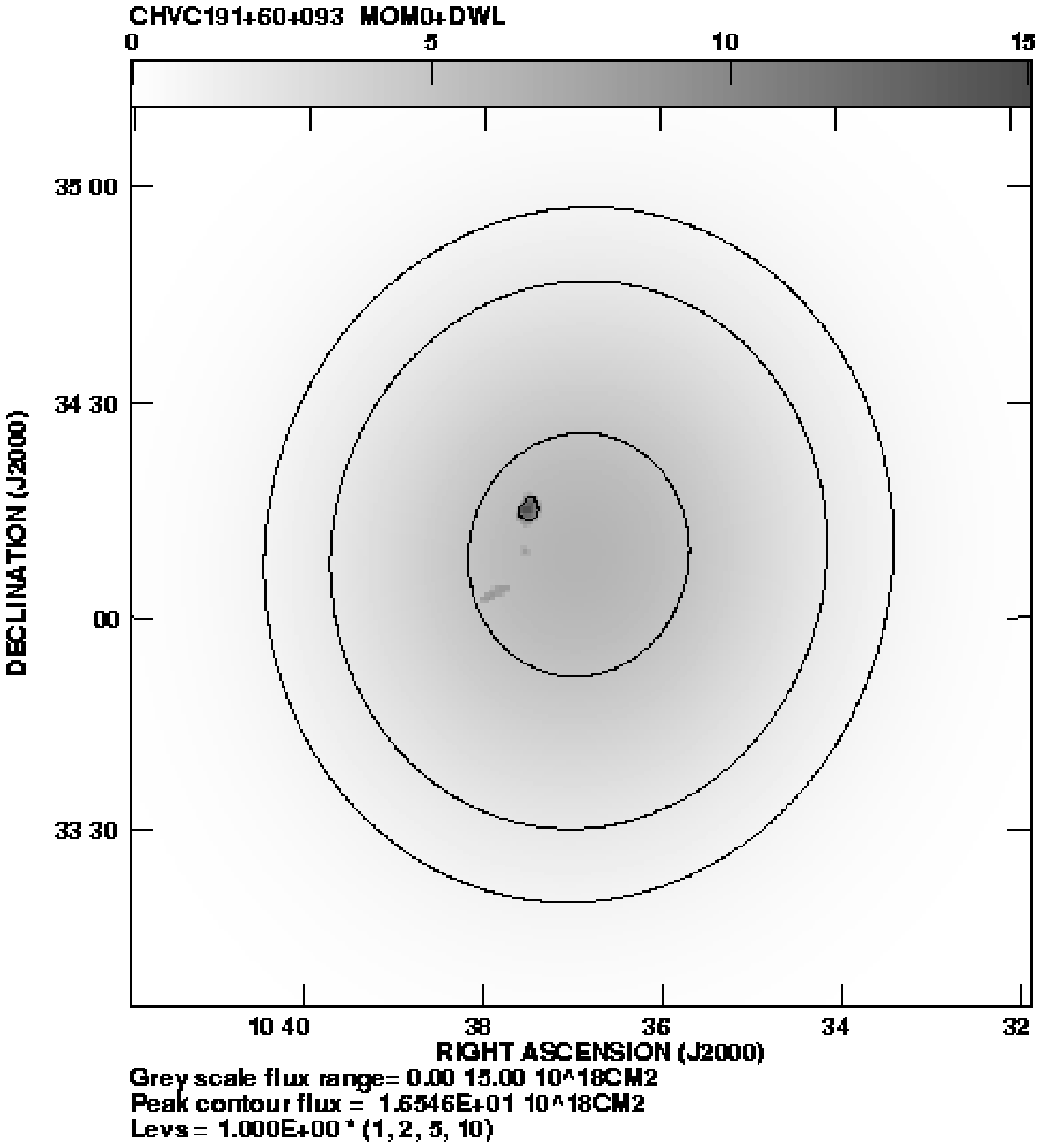}}
\hfill
\parbox[b]{55mm}{
\caption{ Column density distribution of \hi in
  CHVC\,191+60+093 at 2~arcmin resolution reconstructed using LDS and WSRT
  data. Contours are drawn at 1, 2, 5, and 10 
  $\times10^{18}$ cm$^{-2}$ and a linear grey--scale extends from $0$
  to 15$\times10^{18}$ cm$^{-2}$.
 } \label{fig:h191m}}
\end{figure*} 

\begin{figure*}
\resizebox{12cm}{!}{\includegraphics{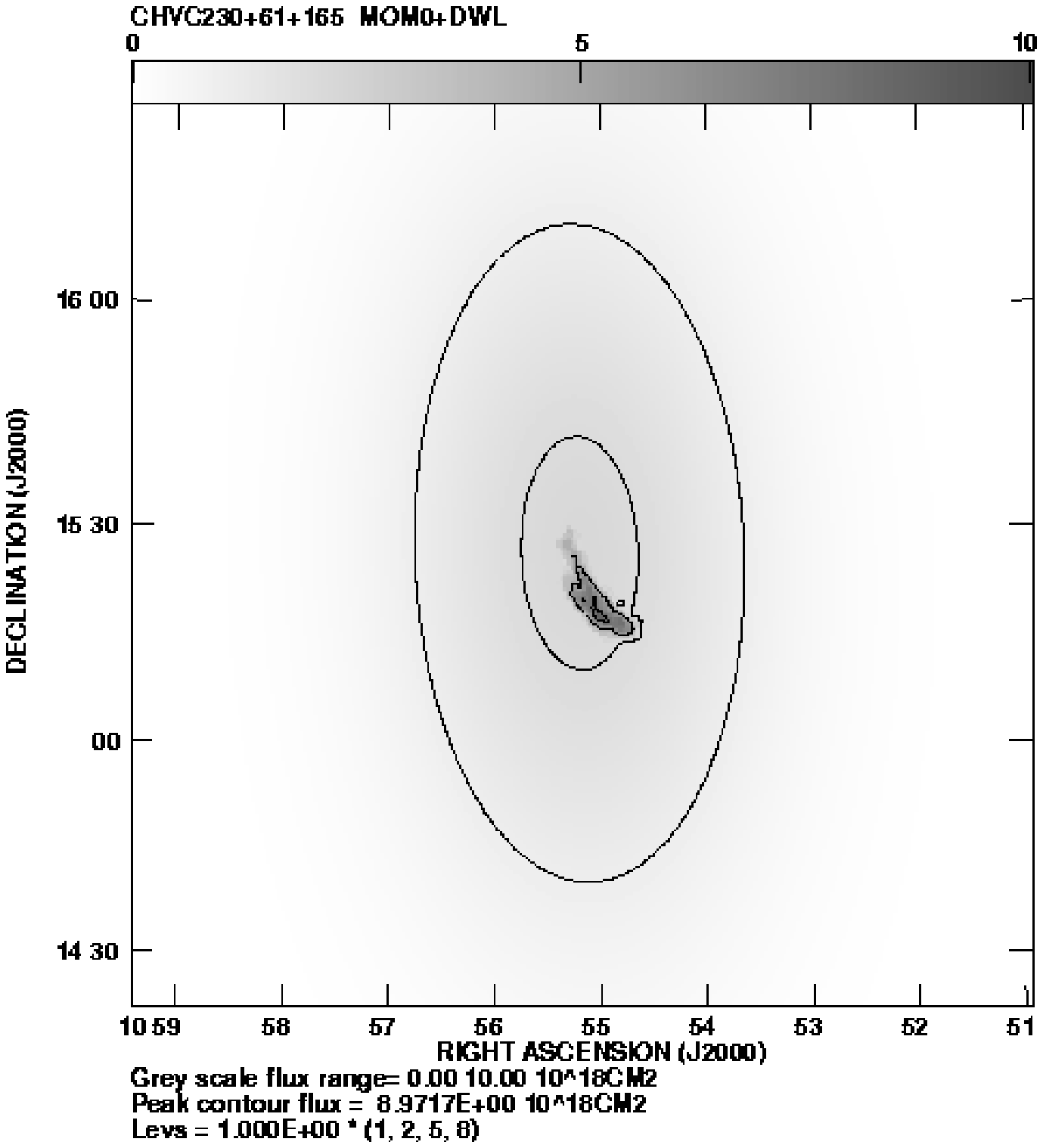}}
\hfill
\parbox[b]{55mm}{
\caption{ Column density distribution of \hi in
  CHVC\,230+61+165 at 2~arcmin resolution reconstructed from LDS and WSRT
  data. Contours are drawn at 1, 2, 5, and 8 
  $\times10^{18}$ cm$^{-2}$ and a linear grey--scale extends from $0$
  to 10$\times10^{18}$ cm$^{-2}$.
 } \label{fig:h230m}}
\end{figure*} 

The two final sources in our sample are both quite faint and primarily
diffuse, albeit small in angular size. The WSRT data at 2~arcmin
resolution are shown in Fig.~\ref{fig:h191o}. Low--brightness clumps of
5 to 10~arcmin extent are detected in both these cases. Peak
brightnesses of less than 0.5~K are seen, coupled with moderately
narrow linewidths of some 6~km~s$^{-1}$ FWHM. The Northern clump of
CHVC\,191+60+093 has a moderately strong velocity gradient along PA
$=0^\circ$ extending from 87 \kms\, to 95~km~s$^{-1}$ over 5~arcmin.
The velocity field of CHVC\,230+61+165 shows no clear systematic trend.
The composite reconstructions shown in Fig.~\ref{fig:h191m} and
Fig.~\ref{fig:h230m} indicate peak column densities of only about
$10^{19}$~cm$^{-2}$, embedded in a diffuse component of only 2--5
$\times10^{18}$~cm$^{-2}$.

\section{Discussion}
\label{sec:disc}

\subsection{Overall morphology and kinematics}

The six compact, isolated high--velocity clouds which we have imaged
with moderately high spatial resolution span a wide range of source
morphologies, but nevertheless share several noteworthy properties. In
each case we detect a number of relatively compact clumps which range
in size from a few arcmin to perhaps 20~arcmin in extent. These are
local enhancements in the column density distribution with peak column
densities in the range $10^{19}$ to $10^{21}$~cm$^{-2}$. Each clump is
characterized by its own line--of--sight velocity and relatively narrow
line widths at 60~arcsec resolution in the range of 2 to 15~km~s$^{-1}$
FWHM. At the narrow linewidth extreme, very stringent limits on the \hi
kinetic temperature are implied which we discuss below. But even
linewidths as large as 15~km~s$^{-1}$ FWHM are significantly less than
the 21~km~s$^{-1}$ FWHM which corresponds to the 8000~K equilibrium
temperature of the Warm Neutral Medium (WNM). Based on these modest
linewidths we can already conclude that the \hi in the CHVC clumps must
be in the form of the Cool Neutral Medium (CNM) with typical
equilibrium temperatures in the range of 50 to 200~K (e.g. Wolfire et
al.  \cite{wolf95a}, \cite{wolf95b}).  Each clump has a smoothly
varying line--of--sight velocity across its extent, often in the form
of a systematic gradient along the major axis of an elliptical
distribution. We will address the best--resolved cases of velocity
gradients below.

Most of the imaged CHVC systems are composed of multiple
clumps. Since each clump has a distinct velocity, the total CHVC
linewidth is determined primarily by the velocity spread between the
clumps, and to a lesser degree by systematic velocity gradients, rather
than by the intrinsic linewidth of a single entity.  There are a few
cases where there appears to be partial line--of--sight overlap of two
distinct clumps, leading to a doubly--peaked emission profile within a
single 60~arcsec beam.  In the most extreme case studied so far,
CHVC\,115+13$-$275, the inter--clump velocity spread amounts to some
70~km~s$^{-1}$. This case will also be addressed in more detail below.

In addition to the clumps, we see evidence, in all cases studied, for a
substantial diffuse halo component of \hi emission. The clumps account
for as little as 1\% to as much as 55\% of the total \hi emission which
was detected in the single--dish observations, although a typical flux
fraction is about 40\%, as was also found previously by Wakker and
Schwarz (\cite{wakk91b}) for the two CHVCs which they imaged. The halos have
typical spatial FWHM of about 1~degree and \hi column densities within
the inner 30~arcmin between 2$\times10^{18}$ and
2$\times10^{19}$~cm$^{-2}$. In Table~\ref{tab:properties} we also include
an indication of the surface area of the high--colum--density cores
relative to the diffuse halos. This was defined as the ratio of the
area with column density exceeding 5$\times10^{18}$cm$^{-2}$ in the
WSRT-only images to the source area determined in the LDS. These values
vary over a wide range, but are typically lower by a factor of about
two from the fractional flux, with a mean of 15\%.  

The data currently
available suggest a characteristic two--phase structure. The diffuse
morphology of the halo is consistent with a component of WNM shielding
the embedded clumps of CNM.  Thermodynamic modeling of the \hi (which we
address in the next subsection) actually requires such a nested
geometry of the warm and cool \hi phases when the entire structure is
immersed in an isotropic radiation field. Comparison of the \hi
emission spectra observed interferometrically with those seen in total
power supports this conjecture, since the diffuse components generally
have linewidths exceeding the 21~km~s$^{-1}$ FWHM thermal linewidth of
an 8000~K gas.  This remains a difficult issue to address in full with
the limited data currently in hand. High--resolution, total--power
imaging with the up--graded Arecibo telescope would allow direct
assessment of the degree of sub--structure as well as the local
lineshape of the halo component.  This could enable unambiguous
determination of the nature of the halo gas.

\subsection{Narrow linewidths and opaque cores in CHVC\,125+41$-$207}

A tight correlation of \hi emission brightness temperature with \hi
opacity has been established for both the Galaxy and for M31 (Braun and
Walterbos \cite{brau92}); a comparable correlation pertains for the
resolved high--brightness \hi structures seen in many nearby galaxy
disks (Braun \cite{brau97}, \cite{brau98}) which have a typical size of
about 150~pc and a velocity FWHM of less than about 5~km~s$^{-1}$. The
high brightness temperatures seen in CHVC\,125+41$-$207, amounting to
about 75~K, are therefore already an indication that relatively high
\hi opacities are likely to be present. In this case a lower limit to
the \hi opacity in the bright cores can be inferred by comparing the
kinetic temperature which follows from the linewidth with the emission
brightness temperature. As illustrated in Fig.~\ref{fig:h125o}, the
linewidth in the cores of clump~A is completely unresolved with the
available 2.47~km~s$^{-1}$ resolution. An upper limit to the intrinsic
linewidth of about 2~km~s$^{-1}$ FWHM corresponds to an upper limit on
the kinetic temperature of 85~K. For an isothermal gas the \hi
brightness temperature, $T_{\rm B}$, is related to the spin
temperature, $T_{\rm S}$, by $T_{\rm B}=T_{\rm S}(1-e^{-\tau}$);
assuming equality of the kinetic and spin temperatures allows a lower
limit on the opacity to be derived of $\tau\ge2$. This must be regarded
as a firm lower limit since any line broadening by turbulence, which
the linewidth observed already constrains to be of order 1~km~s$^{-1}$
or less, would only serve to bring the implied kinetic temperature even
closer to the observed brightness temperature, yielding a
higher $\tau$. Similarly, excitation effects which might suppress the
\hi spin temperature with respect to the kinetic temperature would only
serve to exasperate this problem. (The interesting possibilities for
constraining ISM physics with this remarkable source have prompted us
to request new WSRT observations with substantially higher velocity
resolution.)

Plausible lower limits to the actual \hi column density toward the two
compact cores in clump~A of CHVC\,125+41$-$207 are therefore about
$10^{21}$~cm$^{-2}$ (where the apparent column is some
$4.5\times10^{20}$~cm$^{-2}$). The angular size of these features can
be determined approximately from the apparent column density
distribution shown in Fig.~\ref{fig:h125o}. Mean spatial FWHM extents of about
90~arcsec are observed, although their flat--topped appearance suggests
that opacity effects may already be partially responsible for this
measured size. The FWHM of the actual column density distribution is
likely to be somewhat smaller. These compact cores are seen in
projection against more extended regions with apparent column
densities of 1 to 2$\times10^{20}$~cm$^{-2}$.

The thermodynamics of \hi in a variety of astrophysical settings has
been studied by Wolfire et al. (\cite{wolf95a,wolf95b}). A
stable cool phase of \hi is expected when both $(i)$ a sufficient
column of shielding gas is present, and $(ii)$ the thermal pressure is
sufficiently high. Equilibrium kinetic temperatures of 85~K are
predicted to occur over a range of volume densities, depending on the
dust--to--gas ratio, on the gas--phase metallicity, and on the
intensity of the interstellar radiation field. These dependancies are
illustrated in Figs.~5--7 of Wolfire et al. (\cite{wolf95a}).  Since
the thermal pressure in the mid--plane of our Galaxy in the solar
neighborhood is fairly well--constrained to be about $P/{\rm
  k}=2000$~cm$^{-3}$ K and this pressure is believed to decline
substantially with height above the plane, an upper limit to the volume
density of some 24~cm$^{-3}$ can be assigned to \hi with 85~K kinetic
temperature.

Assuming that the opaque cores in CHVC\,125+41$-$207 have a roughly
spherical geometry, which is indeed suggested by their projected
spatial appearance and that the volume filling factor of cool \hi is
unity, then the column density, angular size, and volume density can be
used to estimate the distance to the source, according to $D=N_{\rm
  H}/(n_{\rm H}\theta)$. This provides a lower limit to the distance to
CHVC\,125+41$-$207 of 31~kpc. Of course at a distance of 30~kpc, the
ambient thermal pressure is likely to have declined by more than an
order of magnitude from the Galaxy mid--plane value (e.g. Wolfire et
al.  \cite{wolf95b}), leading to a corresponding linear increase in the
distance estimate.  At a potential distance of few 100~kpc, it becomes
important to reconsider the ionization and heating conditions that
would apply.

\begin{figure*}
\resizebox{12cm}{!}{\includegraphics{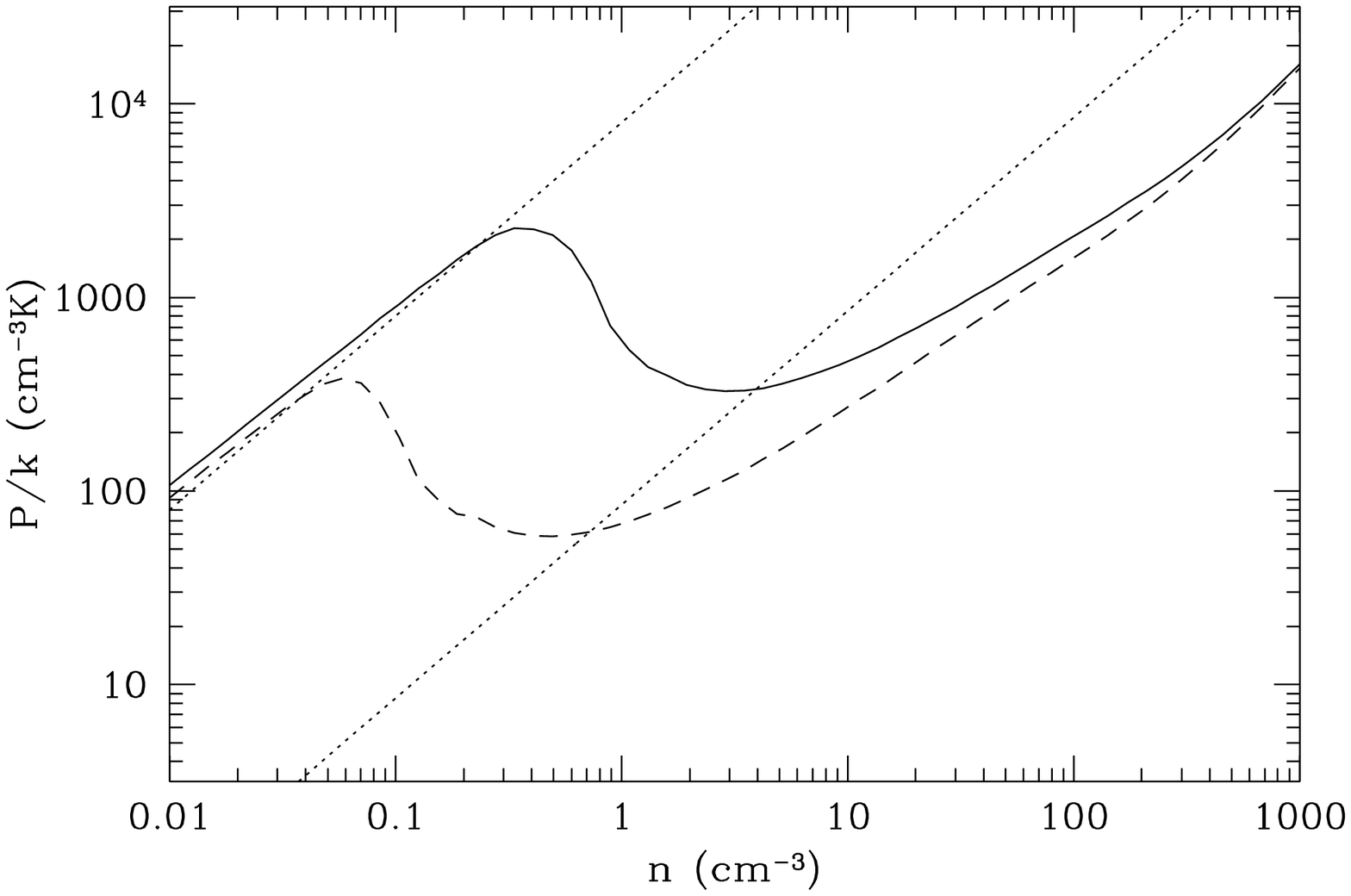}}
\hfill
\parbox[b]{55mm}{
\caption{ Equilibrium temperaturecurves for \hi in an intergalactic
  radiation field at a metallicity of 0.1 solar, a dust--to--gas mass
  ratio of 0.1 times the solar neighborhood value and two values of the
  neutral shielding column density: $10^{19}$~cm$^{-2}$ indicated by
  the solid line and $10^{20}$~cm$^{-2}$ indicated by the
  dashed line. The 85~K kinetic temperature of the opaque cores in
  CHVC\,125+41$-$207 is indicated by the dotted line, as is the
  8000~K typical temperature of the WNM. 
 } \label{fig:phasep}}
\end{figure*}

Specific calculations of equilibrium \hi conditions within the Local
Group environment were made available to us by Wolfire, Sternberg,
Hollenbach, and McKee (private communication) for two bracketing values
of the neutral shielding column density, namely 1 and
10$\times10^{19}$~cm$^{-2}$, a metallicity of 0.1 solar (which we
discuss further below), and a dust--to--gas mass ratio of 0.1 times the
solar neighborhood value. Equilibrium volume densities at $T_{\rm k}=85$~K
of 3.5 and 0.65~cm$^{-3}$ are found for shielding columns of 1 and
10$\times10^{19}$~cm$^{-2}$, respectively, as can be seen in
Fig.~\ref{fig:phasep}. The corresponding distances are in the range of
210 to 1100~kpc. Comparison with Fig.~\ref{fig:h125o} suggests that
shielding columns in the higher part of this range are most relevant
for the two opaque cores under consideration, suggesting a distance in
the range 0.5 to 1~Mpc.

\

We stress that the resulting distance estimate is not dependent on
details of the thermodynamic calculation, but only on the assumed
thermal pressure in the source environment.  Placing CHVC\,125+41$-$207
at a distance of, say, 3~kpc would require assigning it a volume
density of some 250~cm$^{-3}$. The well--constrained temperature of
$T_{\rm k}=85$~K, would then imply a thermal pressure, $P/{\rm
  k}=21000$~cm$^{-3}$ K, a value vastly exceeding those observed in the
local ISM. One might argue that a high pressure may be related to the
high speed re--entry of the object into the Galaxy given its
line--of--sight velocity in the Local Standard of Rest frame of
$-$200~km~s$^{-1}$. It is difficult to reconcile such a scenario with
one of the narrowest \hi emission lines ever detected, showing no
measurable component of turbulent line broadening. A moderately low
thermal pressure, on the other hand, amounting to $P/{\rm k} \sim 100$
cm$^{-3}$ K for this high--latitude, non--turbulent source, seems
plausible on rather general grounds. A source distance in the range 0.5
to 1~Mpc follows directly. In this case a consistent solution for the
heating and ionization of the source is also in hand, as outlined
above.

Unfortunately, this method of distance determination can not be applied
more generally in our current sub--sample. None of the cores in the
other CHVCs we have imaged has a comparably high opacity. The typical
peak brightness seen in other cores amounts to only about 2~K,
suggesting opacities of about 2\%, for $T_{\rm k}\sim100$~K. With such
low opacity, the likelihood of encountering unit filling factor of the
gas along the line-of-sight becomes vanishingly small, so that the
comparison of column and volume densities gives only a very weak lower
limit to the source distance. Applying the same method blindly to the
other cores in our current sample and assuming a volume density of 
1~cm$^{-3}$ throughout, returns lower limits to the distance which vary
substantially within an individual CHVC as shown in
Table~\ref{tab:properties}. The method is particularly suspect for
the low brightness cores (0.5~K peak) of CHVC\,191+60+093 and
CHVC\,230+61+165, where lower limits of 7 and 4 kpc are
found. Elsewhere, lower limits to the distance in the range
40 to 130~kpc are implied.

\subsection{Metal abundance in CHVC\,125+41$-$207}

The source CHVC\,125+41$-$207 is one of the few compact HVCs for which it
has been possible to measure \Mg absorption against a background UV
source. Bowen and Blades (\cite{bowe93}) detected unsaturated \Mg
absorption at a corrected velocity of $v_{\rm LSR}=-209$ \kms\, (Bowen
et al. \cite{bowe95}) toward the Seyfert galaxy Mrk~205 at
$(\alpha,\delta)=(12^{\rm h}21^{\rm m}44\fs12,\,
75^\circ18^\prime38\farcs25)$.  They derive a \Mg column density of
$0.7\pm0.1\times10^{13}$~cm$^{-2}$ for this component.  This line of
sight passes just to the South--East of the high $N_{\rm HI}$ portion
of clump~B shown in Fig.~\ref{fig:h125m}. The column density we infer
in this direction from our composite image is
$5\pm1\times10^{18}$~cm$^{-2}$. Wakker and Van Woerden (\cite{wakk97})
quote a value of $3.0\pm0.7\times10^{18}$~cm$^{-2}$ in a nearby
direction.  Analysis of the recent Jodrell Bank \hi data (de~Vries et
al.  \cite{devr99}) would provide an improved estimate. Jenkins et
al.  (\cite{jenk86}) provide extensive data for the gas--phase
abundance of \Mg relative to \hi in the solar neighborhood. They derive
relative gas--phase abundances of log($N_{\rm MgII}/N_{\rm HI})=-4.7$
and $-$5.0, respectively, for low-- and high--volume density lines of
sight. The relative abundance we infer, log($N_{\rm MgII}/N_{\rm
  HI})=-5.8$, is about a factor of 0.07 down from the local ISM value
appropriate for low volume densities.  Passing as it does through only
the outer reaches of the atomic halo of CHVC\,125+41$-$207, this seems
the most appropriate for comparison.

We have taken care to refer to comparable measurements of
gas--phase metal abundance in comparing the CHVC\,125+41$-$207 data with
that of the solar neighborhood. The local ISM value is believed to be
substantially depleted by deposition on dust grains from its total
value, log($N_{\rm MgII}/N_{\rm HI})=-4.4$. If the dust abundance
of CHVC\,125+41$-$207 is less than that in the solar neighborhood, which is
quite likely in most formation scenarios, then dust depletion would
presumably be less effective and the metal abundance might then lie
somewhere in the range 0.04 to 0.07 relative to the solar value.

\subsection{Localized line splitting}

An interesting kinematic phenomenon to emerge from these high--resolution 
observations of the compact high--velocity clouds is the occurrence of 
several spatially
localized regions which display splitting of the \hi emission profiles
without a related change in the velocity of the centroid of the
profile. An example of the phenomenon is given by the series of
spectra shown in Fig.~\ref{fig:h125s}. Five such regions
were detected, two in each of CHVC\,069+04$-$223 and CHVC\,125+41$-$207, and
one in CHVC\,204+30+075. Each of these cases is associated with a
relatively high column density, in excess of about
$10^{20}$~cm$^{-2}$.  These regions are quite distinct from those
instances of enhanced linewidth which are the result of localized
line--of--sight overlap of extended features having different
velocities.  That phenomenon is also observed on several occasions (as
noted above) but it can typically be recognized as such by the
presence of two distinct velocity systems which are each spatially
extended and have uncorrelated morphology.

The angular extent of these split--line regions is confined to 1 or
2~arcmin in all cases. The degree of velocity splitting varies from
only about 5~km~s$^{-1}$ in CHVC\,125+41$-$207, to 10~km~s$^{-1}$ in
CHVC\,069+04$-$223, and to as high as 20~km~s$^{-1}$ in CHVC\,204+30+075.

Small depressions in the line core of Galactic \hi emission features
have often been interpreted as evidence for so--called self--absorption
by the cool opaque core of a warmer semi--opaque atomic structure. Such
an interpretation does not seem appropriate for the phenomena under
discussion, since the line strength between the peaks declines to a
rather low brightness temperature. The spatially localized and
symmetric distribution of line shapes seen in Fig.~\ref{fig:h125s}
seems more suggestive of an organized inflow or outflow with a
substantial degree of spherical symmetry. At a distance of 700~kpc,
these regions would have linear sizes of 300~pc, making them comparable
to some of the low luminosity super--shells detected in nearby galaxies
which are powered by the supernovae and stellar winds of a stellar
association (e.g.  Mashchenko et al. \cite{mash99}).

\subsection{Velocity gradients}

A common kinematic pattern observed within the CHVC clumps which we
have imaged is a velocity gradient oriented along the major axis of a
roughly elliptical distribution. Specific examples of such gradients
were noted above for several of the objects imaged. The observed
magnitudes of the velocity gradients vary from 0.5 to
2~km~s$^{-1}$\,arcmin$^{-1}$. The cases which are best resolved are
clump~A of CHVC\,069+04$-$223 and clumps A and B of CHVC\,204+30+075, each
of which is about 20~arcmin in angular size. This same pattern was also
seen in the two CHVCs imaged by Wakker and Schwarz (\cite{wakk91b}),
namely CHVC\,114$-$10$-$430 and CHVC\,111$-$06$-$466.

Since these well--resolved examples of systematic gradients in the
velocity fields are reminiscent in form and amplitude of the ``spider''
diagrams seen for \hi distributions in some dwarf galaxies, we carried
out standard tilted--ring fits to assess the extent to which they could
also be modeled by rotation in a flattened disk system. In the usual
way, we began by allowing all kinematic parameters to vary freely from
ring to ring, and then solved sequentially for the best--fitting
kinematic center, for the systemic velocity, for the position angle of
the receding line--of--nodes, and, finally, for the kinematic
inclination. Holding all of these best--fitting parameters fixed, we
then fit only for the rotation velocity of each ring.

\begin{figure*}
\resizebox{12cm}{!}{\includegraphics{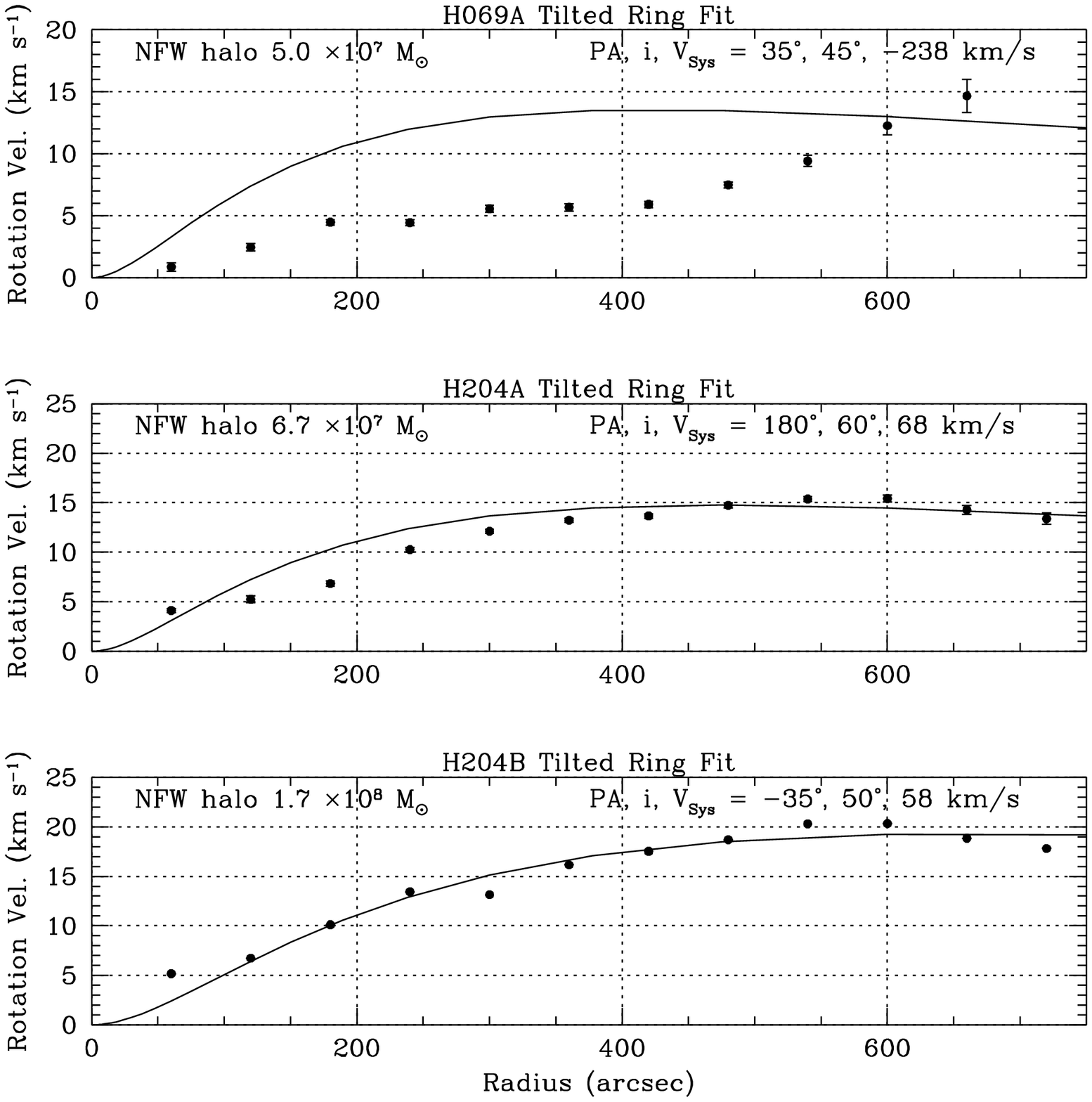}}
\hfill
\parbox[b]{55mm}{
\caption{ Derived rotation velocities in three elliptical cores,
  CHVC\,069+04$-$223A and CHVC\,204+30+075A and B. The best-fitting
 position angle, inclination, and systemic velocity are indicated at the
 top right of each panel. These were subsequently held fixed in deriving the
 rotation velocity as function of radius. Solid lines are the rotation
 curves of NFW cold dark matter halos of the indicated mass.
 } \label{fig:kinenfw}}
\end{figure*} 

Robust solutions for circular rotation were found in all three cases
and these are shown in Fig.~\ref{fig:kinenfw}. The fits display slowly
rising rotation velocity with radius; the rise is continuous to some
15~km~s$^{-1}$ in the case of CHVC\,069+04$-$223A, and flattens out at
500 to 600~arcsec to values of 15 and 20~km~s$^{-1}$ for
CHVC\,204+30+075A and B. An estimate of the contained dynamical mass is
given by $M_{\rm dyn}=Rv^2/G=2.3\times 10^5R_{\rm kpc}v_{\rm km/s}^2$,
while the mass of gas, $M_{\rm gas}=1.4~M_{\rm HI}=3.2\times
10^5S~D_{\rm Mpc}^2$, where $S$ is the integrated \hi flux in units of
Jy\,km~s$^{-1}$ and there is an assumed contribution by Helium of 40\%
by mass. At an assumed distance of 0.7~Mpc, these three clumps have
$M_{\rm dyn}=10^{8.1}$, $10^{8.1}$, and $10^{8.3}$ M$_\odot$, $M_{\rm
  gas}=10^{7.1}$, $10^{6.5}$, and $10^{6.9}$ M$_\odot$, and
dark--to--visible mass ratios of $\Gamma=10$, 36, and 29, respectively.
The derived value of $\Gamma$ scales with the assumed distance as
$1/D$, given the dependencies noted above.

In addition to this crude estimate of dynamical mass, we can compare
our derived rotation curves with those expected for a cold dark matter
halo as parameterized by Navarro, Frenk, and White (NFW,
\cite{nava97}).  These authors find that such halos tend to have a
``universal'' shape in their extensive numerical simulations, so that
the density profile and resulting rotation velocity are determined
simply from the halo mass. We have calculated halo properties using the
(web--retrieved 19/09/99 version) code of Navarro, Frenk, and White
(\cite{nava97}) assuming $\Omega_0$=0.3, $\lambda_0$=0.7,
H$_0$=~60~km~s$^{-1}$~Mpc$^{-1}$ and a cluster abundance normalized
fluctuation spectrum. The early plateau and subsequent rise in the
derived rotation velocity of CHVC\,069+04$-$223A is not well-fit by an
NFW halo. For comparison, the expected rotation curve of a $10^{7.7}$
M$_\odot$ (within the virial radius of 3.4~kpc) NFW halo is overlaid on
the data.  While this curve crudely matches the last few measured
points, the shape at smaller radii deviates substantially. On the other
hand, rather good agreement with the data is found for both
CHVC\,204+30+075A and B with halo masses of $10^{7.8}$, and $10^{8.2}$
M$_\odot$ (within 9.3 and 12.6~kpc) at an assumed distance of 0.7~Mpc.
These rotation curves are overlaid on the data in
Fig.~\ref{fig:kinenfw}. Note how well the shape of these rotation
curves is reproduced by the standard NFW halo profile.

\subsection{CHVC stability}

The very high total linewidth seen for CHVC\,115+13$-$275 was found to
arise from the broad range of distinct line--of--sight velocities of
the ten individual clumps which make up this source. The angular
separation of the clumps is some 30~arcmin, while their velocity
centroids are separated by as much as 70~km~s$^{-1}$. Since this source
is very isolated in position and velocity (see the appropriate panel of
Fig.~1 of Braun and Burton \cite{brau99}) a chance superposition of ten
unrelated components is unlikely.

It is interesting to consider the stability of this collection of
clumps under several scenarios.  If, for instance, this collection were
located at a distance of 5~kpc, the distribution of clumps would have a
linear diameter of 44~pc and would double in size on a timescale of
only $5\times10^5$~yr. Such a short dynamical timescale would imply
that we were witnessing a rather special moment in the evolution of
this source.  If, on the other hand, this source were located at a
distance of 0.7~Mpc and were self--gravitating, the angular radius and
half--velocity width of 15~arcmin and 35~km~s$^{-1}$ could be used to
calculate a dynamical mass from $M_{\rm dyn}=Rv^2/G=2.3\times
10^5R_{\rm kpc}v_{\rm km/s}^2=10^{8.93}$ M$_\odot$.  The corresponding
mass of gas at this distance, $M_{\rm gas}=1.4~M_{\rm HI}=3.2\times
10^5S~D_{\rm Mpc}^2$ is $M_{\rm gas}=10^{7.22}$ M$_\odot$. The
dark--to--visible ratio at this assumed distance is $\Gamma$~=~51,
scaling, as in the previous subsection, with $1/D$.

A comparable analysis can be applied to CHVC\,043$-13-302$, for which
Arecibo imaging data was obtained by Giovanelli (\cite{giov81}).
This object has only two primary core components which extend over
about 40~arcmin, and are separated by 60~km~s$^{-1}$ in velocity, while
the total line flux is 260~Jy-km~s$^{-1}$. Using the half-separations
and line flux yields $M_{\rm dyn}=10^{8.92}$ M$_\odot$, $M_{\rm
  gas}=10^{7.61}$ M$_\odot$ and $\Gamma$~=~20 for $D$~=~0.7~Mpc. Since
only two core components are involved, geometric effects
of orientation are more likely to play a role in this calculation than
in the case of CHVC\,115+13$-$275.

\subsection{Comparison with nearby galaxies}

The \hi imaging obtained here for our CHVC sub--sample reveals striking
similarities between the gas properties of the CHVCs and those of low
mass galaxies. The comparison with the Local Group dwarf
irregulars, Leo~A and Sag\,DIG, studied at high resolution by Young and
Lo (\cite{youn96}, \cite{youn97b}) is particularly apt. Both of these
objects display the same general morphology and kinematics as the
CHVCs: high column density clumps characterized by narrow linewidths
(about 8~km~s$^{-1}$ FWHM) and angular sizes of a few arcmin embedded
in a diffuse halo of lower column density gas with broader linewidths
(about 20~km~s$^{-1}$ FWHM) extending over 10's of arcmin.  Young and
Lo reach the same conclusion presented here regarding the \hi phase
content of the dwarf irregulars, namely that cool condensations of CNM
(with $T_{\rm k}\sim$100~K) are found within WNM envelopes (with
$T_{\rm k}\sim$8000~K). The peak brightness temperatures detected from the
CNM in Leo~A, 73~K, are almost identical to those we see in the compact
clumps of CHVC\,125+41$-$207. 

The velocity fields of both Leo~A and Sag\,DIG are also very similar to
those of the CHVCs. Modest velocity gradients of comparable magnitude
are observed (about 10~km~s$^{-1}$ over 10~arcmin) along one position
angle in addition to a substantial unstructured component. Unlike the
CHVCs, these galaxies contain comparable gas and stellar masses, each
corresponding to about 10$^8$ M$_\odot$ in the case of Leo~A and about
10$^7$ M$_\odot$ in Sag\,DIG.

Young and Lo also comment on a small region of about 2.5~arcmin
diameter within Leo~A which they suggest might be the expansion or
contraction signature of an \hi shell, comparable to the five cases of
line--splitting within CHVCs which we describe above.

The dwarf elliptical companions of M31, NGC~185 and NGC~205 (Young and
Lo \cite{youn97a}) are rather different in their gas properties.  While
both the CNM and WNM components of \hi are seen in these two cases, all
of the neutral gas is concentrated to the central five arcmin (1~kpc)
of each galaxy and the global kinematics are less systematic. The
strong interaction of these objects with M31 (NGC~205 in particular is
on an orbit that is likely to have recently passed through the disk of
M31) and the high mass density of stars (stellar masses about 100 times
greater than the 10$^{5.5}$M$_\odot$ in gas) makes a direct comparison
with the CHVCs less obvious.

An intermediate and less--perturbed example is given by the more
isolated Local Group dwarf spheroidal LGS~3 (Young and Lo
\cite{youn97b}). In this case, the stellar mass is likely only a few
times that of the 10$^{5.8}$ M$_\odot$ in gas. Here the \hi is also
confined to only the inner 5~arcmin (corresponding to 1~kpc), shows
little systematic kinematics, and the cool condensed phase is almost
entirely absent. Although peak \hi column densities of somewhat more
than 10$^{20}$ cm$^{-2}$ are observed over a region of a few arcmin
extent, there is no sign of ongoing star formation.

Extending the comparison to even higher mass systems provides some
further insight into the conditions necessary for allowing the
condensation of CNM clumps in a galactic environment. Several low mass
spirals were observed in the high resolution \hi sample of Braun
(\cite{brau95}, \cite{brau97}), including the nearby SBm galaxy
NGC~2366 and two of the nearest low surface brightness spirals, NGC~247
(of type Sc) and NGC~4236 (SBd). A filamentary network of CNM gas is
seen in all these systems, which accounts for about 80\% of the \hi line
flux from only about 15\% of the surface area within $R_{25}$ (the
radius where the average face-on surface brightness of light in the B
band has declined to 25 mag arsec$^{-2}$).  Just beyond $R_{25}$ the
flux fraction of CNM gas disappears entirely, even though the WNM
continues out to a radius of almost twice $R_{25}$. The result is
that the {\it total\ } flux fraction due to CNM in these galaxies is
about 70\%, substantially more than the 40\% CNM fraction in the CHVCs.
Comparison with the \hi thermodynamics of Wolfire et al.
(\cite{wolf95a}) suggest that the edge of the CNM disk should occur
where the thermal pressure has declined to a value of $P_{\rm min}$, below
which CNM condensation is no longer possible. The numerical value of
$P_{\rm min}$ depends on the dust and gas phase metal abundance, the
radiation field intensity and the shielding column of neutral gas.
Values for $P_{\rm min}$/k of about 200~cm$^{-3}$~K seem indicated under
conditions of moderate metallicity and radiation field-intensity
expected for the low mass spirals above.  Comparison with
Fig.~\ref{fig:phasep} suggests that slightly lower values for
$P_{\rm min}$/k of about 100~cm$^{-3}$~K should be expected under the low
metallicity, low field--intensity conditions that might apply to the
CHVCs.

What pressures might be expected in the mid-plane of a disk--like
self--gravitating distribution of \hi$\!$? The concept of hydrostatic
pressure equilibrium demands that the weight of the overlying medium is
balanced by the sum of the contributions to the gas pressure. This can
be expressed as
\begin{eqnarray}
P_{\rm tot}(z)  & = & P_{\rm B} + P_{\rm CR} + P_{\rm turb} + P_{\rm th} +
\, \dots \\
 & = & \int_z^\infty 4 \pi G \Sigma_{\rm tot}(z^\prime) 
\rho_{\rm gas}(z^\prime)dz^\prime  
\label{eqn:hydro}
\end{eqnarray}
where $\rho_{\rm gas}(z)$ is the gas volume density at height $z$ from the
mid-plane. In this form, $\Sigma_{\rm tot}$ refers to the mass surface
density between $z'=0$ to $z'=z$, not between $z'=-z$ to $z'=z$. The
total pressure is composed of magnetic, cosmic ray, and turbulent
components in addition to the thermal component which is of relevance
to the calculation of the HI kinetic temperature. In a high mass
galactic disk, the thermal pressure might amount to only about 1/4 of
the total pressure, while in a low mass CHVC with few if any sources of
internal turbulence and energetic particles, the thermal pressure might
be the dominant component. If for simplicity we assume a constant gas
density, $\rho_{\rm gas}$, for $-s/2~<~z~<~+s/2$ for some full disk
thickness, $s$, then 
$\Sigma_{\rm tot}(z)=(\rho_{\rm gas}+\rho_{\rm stars}+\rho_{\rm DM})\times z$
for $z$ less than $s/2$. For the moment we will 
neglect contributions to the mass surface density by stars or dark matter,
which if present would lead to proportionally higher pressures.
Equation \ref{eqn:hydro} then becomes simply
\begin{equation}
P_{\rm tot}(0) = \pi G \rho_{\rm gas}^2 s^2 / 2 = \pi G (\mu m_{\rm H} 
N_{\rm H})^2/2.
\label{eqn:press}
\end{equation}
Inserting numerical values into eqn. \ref{eqn:press} yields
\begin{equation}
P_{\rm tot}(0)/{\rm k} = 41 (N_{\rm H}/10^{20})^2 \ \ {\rm cm^{-3}~K}.
\label{eqn:n20}
\end{equation}
Thus, whenever the face--on \hi column density exceeds about
10$^{20}$ cm$^{-2}$ in a disk--like system we can expect to see the cool
condensations of the CNM. Given the quadratic dependence on column
density in eqns. \ref{eqn:hydro}--\ref{eqn:n20}, we might expect the
transition from purely WNM to a CNM/WNM mix to be quite abrupt. This
expectation is borne out quite dramatically at the edge of the CNM
disks in spiral galaxies which indeed occurs very near a face--on \hi
column density of 1--2$\times$10$^{20}$~cm$^{-2}$. If a significant
contribution to the mass surface density within one gas scale height
were due to stars or dark matter then the minimum column density needed
for CNM condensation would decrease. For the CHVCs, our current data
suggest a transition column density nearer 10$^{19}$ cm$^{-2}$, rather
than the 10$^{20}$ cm$^{-2}$ seen in nearby spirals. This may be an
indication that the mass surface density of dark matter is already a
significant contribution.

\begin{figure*}
\resizebox{12cm}{!}{\includegraphics{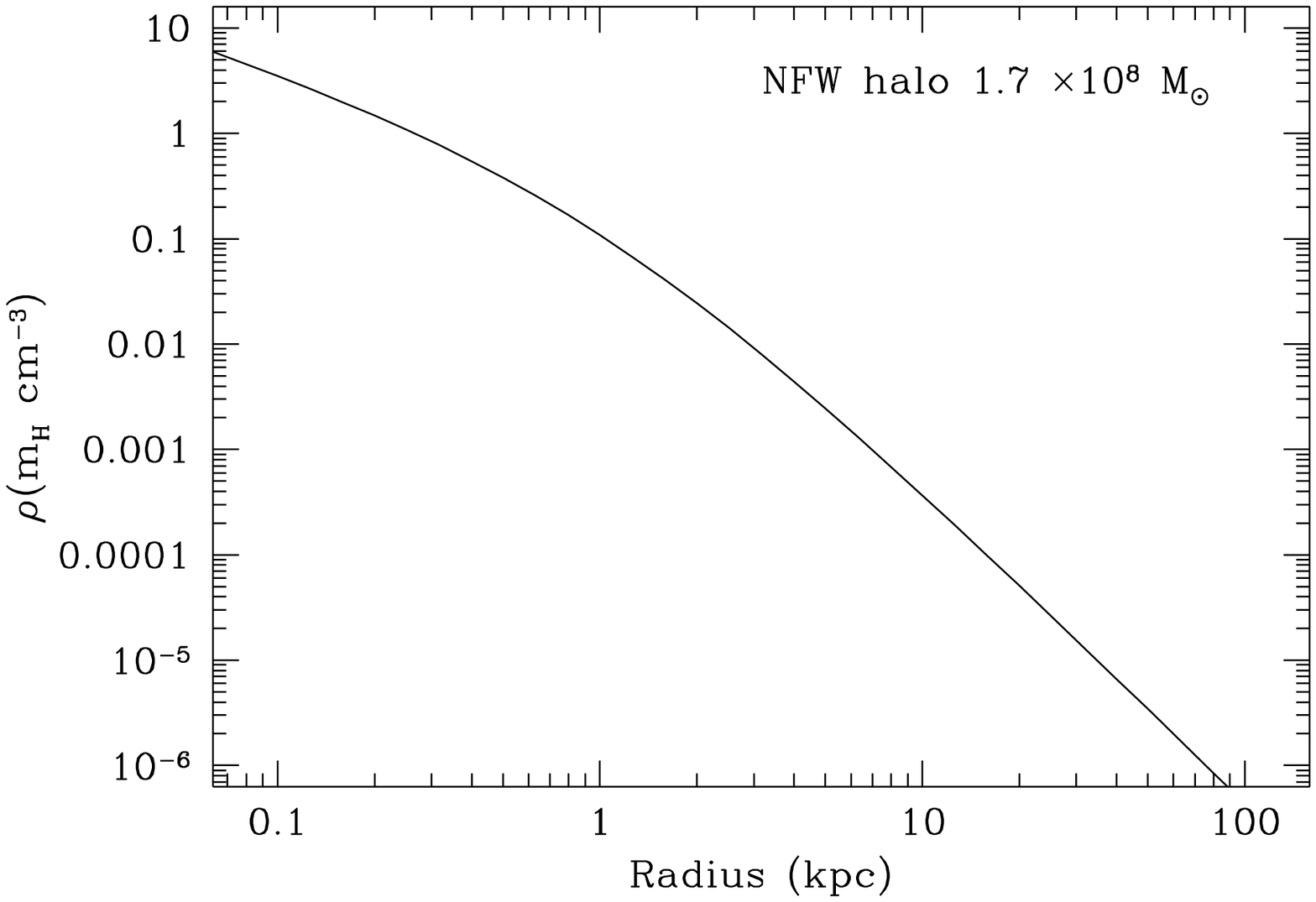}}
\hfill
\parbox[b]{55mm}{
\caption{ Mass volume density, in units of hydrogen nuclei per cubic
  centimeter, as a function of radius for an NFW cold dark matter halo
 of mass $10^{8.2}$ M$_\odot$ within the 12.6~kpc virial radius.
 } \label{fig:rhoplot}}
\end{figure*}

We can assess this possibility by considering the mass volume density
of the implied dark halos. The density distribution of the $10^{8.2}$
M$_\odot$ NFW halo which gave a good fit to the derived rotation
velocity of the core in CHVC\,204+30+075B (Fig.~\ref{fig:kinenfw}) is
shown in Fig.~\ref{fig:rhoplot}. At a thermal pressure of
$P/{\rm k}~=~100$~cm$^{-3}$~K, the 8000~K WNM halos of the CNM cores 
will have a volume density of about 0.01~cm$^{-3}$. Comparison with
Fig.~\ref{fig:rhoplot} shows that the dark halo will provide a
dominant contribution to the mass density out to several kpc. Within
the central region, of perhaps 2~kpc (or 10~arcmin at $D~=~0.7$~Mpc) in
diameter, the mass density is enhanced by more than a factor of 10 over
that of the gas alone. In consequence, we can expect to achieve values
of $P_{\rm min}$/k of about 100~cm$^{-3}$~K in the CHVCs whenever the \hi
column densities exceed a few times 10$^{19}$~cm$^{-2}$.

\section{Summary}
\label{sec:summ}

We present some of the first high--resolution \hi imaging of a
sub--sample of the compact high--velocity clouds (CHVCs). Previously we
had demonstrated that the spatial and kinematic properties of the CHVCs
were strongly suggestive of a Local Group origin (Braun \& Burton
\cite{brau99}), although the typical distance could not be
well--constrained from the low resolution (0\fdg5) data then in hand.

Our imaging has revealed that these objects have a characteristic
morphology, consisting of one or more quiescent, low--dispersion compact cores 
embedded in a
diffuse, warmer halo. This is consistent with what was seen previously by
Wakker \& Schwarz (\cite{wakk91b}) for the two CHVCs which they imaged.
The compact cores can be unambiguously identified with the cool neutral
medium (CNM) of condensed atomic hydrogen, since their linewidths are
in all cases significantly narrower than even the thermal linewidth of
the warm neutral medium (WNM). In one object we detect some of the
narrowest \hi emission lines yet seen, with an intrinsic FWHM 
of 2~km~s$^{-1}$. The halo linewidths, while difficult to measure
directly, appear broad enough to be consistent with a WNM origin. Such
a nested geometry of CNM and WNM is in agreement with the expectation
from thermodynamics since a significant neutral shielding column is
required for the stable existence of CNM when exposed to an ionizing
radiation field (e.g. Wolfire et al. \cite{wolf95a,wolf95b}). The flux
fraction in compact cores varies between about 1\% and 50\% in our CHVC
sub--sample, but is typically 40\%, as also seen in the two cases
studied by Wakker and Schwarz (\cite{wakk91b}). The surface covering
factor of the moderate column density core emission relative to the
diffuse halo is usually a factor two lower, namely about 15\%.

High--resolution imaging of the \hi column density distribution of
CHVC\,125+41$-207$ has permitted an improved determination of the metal
abundance in this object when combined with a previous measurement of
unsaturated \Mg absorption (Bowen \& Blades \cite{bowe93}, Bowen et al.
\cite{bowe95}) of between 0.04 and 0.07 solar.

The current program has also allowed meaningful distance estimates for
CHVCs to
be made for the first time.  Distances in the range 0.5 to 1 Mpc follow
from two independent approaches. The first method utilizes the fact
that we have resolved cool, opaque knots of \hi within one of the six
imaged systems (CHVC\,125+41$-207$). By equating the observed angular
size of these knots with the line--of--sight depth, we derive a
distance from a comparison of the column and volume
densities. Estimates 
of the column density follow directly from our data, while
specific thermodynamic calculations for volume densities under the
relevant physical conditions were made available to us by Wolfire,
Sternberg, Hollenbach, and McKee (private communication). The second
approach allows only a combination of source distance and dark matter
fraction to be derived, since it relies on a comparison of dynamical
with visible mass under the assumption that the system is
gravitationally bound. For this approach, the CHVC\,115+13$-275$ system
proved particularly illuminating, since a collection of some 10 compact
cores is distributed over a 30~arcmin diameter region within a common
diffuse envelope, while having discrete line--of--sight velocities
distributed over a 70~km~s$^{-1}$ range. At an assumed distance,
$D~=~0.7$~Mpc, the dark--to--visible mass ratio is $\Gamma = 51$, with
$\Gamma$ scaling as $1/D$.

A common feature observed in the CHVCs is that the compact cores have
an elliptical shape with a clear velocity gradient aligned with the
major axis. These kinematic data can be well--modeled by rotation in
flattened disk systems.  Rotation curve fits to the best--resolved cores
within CHVC\,069+04$-223$ and CHVC\,204+30+075, yield rather high
dark-to-visible mass ratios of $\Gamma$ = 10--40 at D~=~0.7~Mpc, with
$\Gamma$ scaling, as above, with 1/D. Moreover, the shape of the
rotation curves in two of these three cases is very well described by
that due to a cold dark matter halo as parameterized by Navarro et al. 
(\cite{nava97}).

Well--resolved imaging has also revealed another intriguing aspect of the
CHVCs. Five instances of localized line splitting are detected. These
are areas of 60 to 90 arcsec extent in regions of moderately high
column density in which the \hi emission profile becomes doubly peaked in
velocity, without displaying a shift in the velocity centroid. This
appears to be a signature of localized outflow or inflow. At a distance
of 0.7 Mpc they would correspond to feaures of a few hundred pc in
extent, comparable in size to the stellar--wind--driven bubbles
surrounding young stellar associations. 

Comparison of the CHVC properties with those of nearby low mass
galaxies shows striking similarities, particularly with the Local Group
dwarf irregulars. The required thermal pressure for condensation of the
cool CNM cores, $P_{\rm min}$/k~$\sim$~100~cm$^{-3}$~K, can be achieved in
hydrostatic equilibrium of a self--gravitating disk--like distribution at
the observed shielding \hi column of WNM of about 10$^{19}$~cm$^{-2}$ if
the significant mass density of the dark matter halos is taken into
account within the inner 2~kpc.

Our study of the compact high--velocity--clouds has helped to clarify
the nature of these objects. All of the evidence points to these being
strongly dark--matter--dominated, self--gravitating objects of modest
mass in the Local Group potential at distances between perhaps 0.3 and
1 Mpc with physical dimensions in \hi of about 10~kpc.  Typical
\hi masses of a few times 10$^7$~M$_\odot$ are indicated, with
associated dark masses of some 10$^9$~M$_\odot$. The total number of
such objects associated with the Local Group, while still uncertain,
may be about 200. It seems likely that these are the objects which
recent numerical simulations suggest should still be associated with
massive galaxies in a poor environment at the current epoch (Klypin et
al. \cite{klyp99}, Moore et al. \cite{moor99}).

The star formation history within these objects is
still very much a mystery. High stellar densities, like those seen in
dwarf spheroidal systems in the Local Group, are clearly not associated
with the centroids of the gas distribution. A large spatial separation
of the gas and associated stellar distribution, due for example to
tidal effects, is not expected to be a common circumstance since the
population is dynamically so cold. The peak \hi column densities
currently observed within the CHVC cores only rarely exceed
10$^{20}$ cm$^{-2}$. It is plausible that these values are
insufficient to allow widespread star formation at this time. On the
other hand, finite yet low metallicities have been measured in at least
one source, indicating that some enrichment has clearly taken place in
the past. It is quite possible that associated old stellar populations
with a low surface density will be found in deep searches which can
distinguish such stars from the Galactic foreground.

\begin{acknowledgements} 
  
  We are grateful to M.G. Wolfire, A. Sternberg, D. Hollenbach, and C.F.
  McKee for providing the equilibrium temperature curves for \hi in an
  intergalactic radiation field shown in Fig.~\ref{fig:phasep} as well
  as stimulating correspondence on the ISM in general, and to
  U. Schwarz and H. van Woerden for showing us, in advance of
  publication, the Jodrell Bank data of de Vries et al. (\cite{devr99})
  on the source CHVC\,125+41$-$207.  The Westerbork Synthesis Radio
  Telescope is operated by the Netherlands Foundation for Research in
  Astronomy, under contract with the Netherlands Organization for
  Scientific Research.

\end{acknowledgements}


\end{document}